\documentclass[aps,pre,preprint,superscriptaddress]{revtex4-1} 

\usepackage{amsfonts}
\usepackage{amsmath}
\usepackage{amssymb}
\usepackage{mathrsfs} 
\usepackage{mathrsfs}
\usepackage{color} 
\usepackage{graphicx}

\providecommand\bnabla{\boldsymbol{\nabla}}

\providecommand\bcdot{\boldsymbol{\cdot}}

\providecommand\bv{\mathbf{v}}
\providecommand\bx{\mathbf{x}}

\providecommand\bE{\mathbf{E}}
\providecommand\bB{\mathbf{B}}

\providecommand\be{\mathbf{\hat{e}}}
\providecommand\bn{\mathbf{\hat{n}}}

\providecommand\unit{\boldsymbol{\hat{\imath}}}

\newcommand{\pd}[2]{\frac{\partial #1}{\partial #2}}

\newcommand{\ub}[1]{^{({#1})}}

\begin{document}
\title{Surface-plasmon resonances of arbitrarily shaped nanometallic structures in the small-screening-length limit} 
\author{Ory Schnitzer}
\affiliation{\footnotesize{Department of Mathematics, Imperial College London, London SW7 2AZ, United Kingdom}}
\author{Vincenzo Giannini}
\affiliation{\footnotesize{The Blackett Laboratory, Department of Physics, Imperial College London,  London SW7 2AZ, United Kingdom}}
\author{Stefan A.~Maier}
\affiliation{\footnotesize{The Blackett Laboratory, Department of Physics, Imperial College London,  London SW7 2AZ, United Kingdom}}
\author{Richard V.~Craster}
\affiliation{\footnotesize{Department of Mathematics, Imperial College London, London SW7 2AZ, United Kingdom}}

\begin{abstract}
{According to the hydrodynamic Drude model, surface-plasmon resonances of metallic nanostructures blueshift owing to the nonlocal response of the metal's electron gas. The screening length characterising the nonlocal effect} is often small relative to the overall dimensions of the metallic structure{,} which enables us to derive a coarse-grained nonlocal description using matched asymptotic expansions; a perturbation theory for the blueshifts of arbitrary shaped nanometallic structures is then developed. The effect of nonlocality is not always a perturbation and we present a detailed analysis of the ``bonding'' modes of a dimer of nearly touching nanowires where the leading-order eigenfrequencies and eigenmode distributions are shown to be a renormalisation of those predicted assuming a local metal permittivity. 
\end{abstract}

\maketitle

\section{Introduction}
Metallic nanostructures supporting localised surface-plasmon resonances provide means for channeling electromagnetic energy between far and near optical fields \cite{Maier:07}. Recent interest has focussed on the ultimate limits of plasmon-assisted light confinement \cite{Schuller:10,Ciraci:12,Raza:15b}, particularly in the context of assemblies of 
metallic nanostructures separated by nanometric gaps, such as a dimer of nearly touching particles. Confinement of the electromagnetic field to dimensions comparable to those characterising the solid-state physics, leads to the breakdown of a purely classical description of the surface plasmon in the framework of the macroscopic Maxwell's equations with bulk values of the metal's permittivity.

A first correction, taking electron-electron interactions into account, is the nonlocal description of the material's polarisibility, and an implementation utilising the hydrodynamic Drude model \cite{David:11,Ciraci:13,Raza:13,Christensen:14,Mortensen:14,Raza:15} has, despite not describing the full physical reality of noble metals \cite{Zuloaga:09,Esteban:12,Scholl:12,Marinica:12,Savage:12,Scholl:13,Haus:2014,Toscano:15,Zapata:16}, led to a deeper understanding of fundamental constraints on nanoscale plasmon-assisted light localisation \cite{Ciraci:13b,luo:14,Raza:15b}. 
Specifically, the case of nearly touching nanometallic structures has been studied numerically \cite{Garcia:08,Ciraci:12} and also via approximate analytical solutions based on transformation optics \cite{Fernandez:12,Fernandez:12b,luo:14} and an intuitive local-analogue model \cite{luo:13}. At small separations, the hydrodynamic Drude model predicts that the familiar surface-plasmon redshift of the ``bonding'' modes with vanishing gap separation arrests at separations on the order of the {Thomas--Fermi screening} length. This striking prediction is consistent --- up to the threshold for quantum tunnelling --- with both detailed quantum-mechanical simulations \cite{Stella:13} and experiments of light scattering from a gold sphere in near contact with a gold substrate \cite{Ciraci:12}. 

The notion that nonlocality at small scales is crucial in order to predict the surface-plasmon redshift saturation for the bonding gap modes has rekindled interest in the effects of nonlocality on the optical properties of isolated nanometric particles. Exact analytical solutions of the nonlocal hydrodynamic equations have been obtained and thoroughly analysed for spheres and circular nanowires \cite{Klimov:14,Christensen:14,Raza:15}, and there have also been substantial efforts to implement general  codes able to overcome the challenging task of solving these equations numerically for particles of arbitrary shape \cite{Mcmahon:09,Mcmahon:10,Yan:13}. For isolated particles, surface-plasmon blueshifts predicted by the hydrodynamic Drude model are typically small (excluding tiny subnanometric particles); nevertheless, they have been observed in electron energy-loss spectroscopy experiments of nanoparticles \cite{Raza:13}. Furthermore they may play an important role in optical sensing and plasmon-ruler applications, especially since, contrary to quasistatic eigenfrequencies \cite{Mayergoyz:05}, nonlocal eigenfrequency corrections depend on size \cite{Tserkezis:16}, in addition to shape and mode number.

We have recently put forward a novel theoretical approach for studying surface-plasmon resonances \cite{Schnitzer:15:dimer}, and effects of nonlocality in particular \cite{Schnitzer:16}, based on the paradigm of singular perturbation theory \cite{Hinch:91}. Plasmonic phenomena are often linked with physical and geometric scale disparities, that can be systematically exploited to derive formulae for frequency  eigenvalues and field enhancements. The approach entails scaling arguments, and the divide and conquer strategy of matched asymptotic expansions, where minimal mathematical descriptions of physically distinct regions together constitute a complete asymptotic solution of an otherwise intractable problem. This facilitates a clear physical picture of surface-plasmon resonances and their near-field distributions, and in particular allows us to \emph{a priori} identify and characterise spatial domains wherein nonlocality is important and field intensity is high.

{In Ref.~\onlinecite{Schnitzer:16} we}
considered the plasmonic eigenvalue problem for a generic dimer system, showing via scaling and  asymptotic arguments that, in the near-contact limit, nonlocality acts to renormalise the otherwise singular surface-plasmon redshift of the dimer's bonding modes; when the {nonlocal screening length} is small relative to particle size, the deviation of the electron density from equilibrium is exponentially confined to a narrow boundary layer adjacent to the metal-vacuum interface. The separation of scales between the {boundary-layer thickness} and those on which the near-field and electron-charge distributions vary along the interfaces, led 
 to a coarse-grained local-analogue eigenvalue problem. In the resulting physical picture,  the near-contact redshift saturation, or renormalisation, is effectively captured by {a widening of the gap}. 

Here we perform a detailed asymptotic analysis of the nonlocal hydrodynamic Drude model in the prevalent limit where {the nonlocal screening length}
is small relative to the overall dimensions of the nanometallic structure. {In that limit, and for smooth particles characterised by a single length scale, the local-analogue model of Ref.~\onlinecite{Schnitzer:16} suggests that nonlocality is a small perturbation.} Thus, a main goal of this paper is to develop and demonstrate a perturbation theory --- applicable to nanometallic structures of arbitrary shape --- giving the surface-plasmon blueshifts resulting from nonlocality as some functional of the local-theory eigenmodes. A second goal of this paper is to substantiate the local-analogue model of Ref.~\onlinecite{Schnitzer:16} on the basis of the method of matched asymptotic expansions. Furthermore, we {aim to clarify}
the justification for applying this model in the {near-contact limit of a dimer structure}, where the effect of nonlocality is no longer a small correction, and also in excitation scenarios.

In \S\ref{sec:form} we formulate the nonlocal hydrodynamic Drude model, note the existence of quasistatic plasmonic eigenfrequencies and eigenmodes, and discuss several near- and far-field excitation scenarios. In \S\ref{sec:sphere} we revisit the analytically tractable problem of a metallic nanosphere. In \S\ref{sec:arbitrary} we carry out an \emph{ab initio} asymptotic analysis of the surface-plasmon eigenmodes of arbitrarily shaped structures characterised by a single length scale. 
A detailed analysis of the near-contact limit of a cylindrical dimer is carried out in \S\ref{sec:near}. In \S\ref{sec:macro} we relate the coarse-grained model employed in Ref.~\onlinecite{Schnitzer:16} with that derived in \S\ref{sec:arbitrary}, and discuss its applicability more generally as a uniformly valid model, and in the context of surface-plasmon excitation problems. Concluding remarks are given in \S\ref{sec:conclude}, including a recapitulation of the key results in dimensional form.

\section{Formulation}\label{sec:form}
\subsection{Hydrodynamic Drude model} \label{ssec:hydro}
The hydrodynamic Drude model is derived in, say, \cite{Ciraci:13,Raza:15} and we give a brief recapitulation, assuming for simplicity that the metal is surrounded by vacuum and that metal polarisation arises solely from deviations of the electron density from equilibrium. Accordingly, within the metal, Gauss's law is
\begin{equation} \label{Gauss}
\epsilon_0\bnabla\bcdot\bE = -e(n-n_e),
\end{equation}
where $\bE$ denotes the microscopic electric field, $n$ the electron density and $n_e$ an equilibrium electron density; in vacuum \eqref{Gauss} holds with the right hand side set to zero. The electron density $n$ and hydrodynamic velocity $\bv$ are governed by the continuity and momentum equations
\begin{equation}\label{cont NS}
\pd{n}{t}=-\bnabla\bcdot(n\bv), \quad \left(\pd{}{t}+\bv\bcdot\bnabla\right)\bv = -\frac{e}{m}\left(\bE+\bv\times\bB\right) -\gamma \bv -\beta^2\bnabla\ln n,
\end{equation}
respectively.
The first term on the right hand side of the momentum equation is the Lorentz force ($\bB$ being the magnetic induction and $m$ the effective electron mass), the second is a phenomenological dissipation term ($\gamma$ is a collision frequency), and the third is an electron-pressure term derived from the Thomas--Fermi energy functional; the parameter $\beta$ is a nonlocality parameter, that, for frequencies $\omega\gg\gamma$, is $\beta^2=3\nu_F^2/5$, $\nu_F$ being the Fermi velocity. Assuming that the electrons are confined to the metal domain, the metal-vacuum interfacial conditions read
\begin{equation} \label{bcs}
[\bn\bcdot \bE] =0, \quad [\bn\times \bE]=0, \quad \bn\bcdot \bv=0,
\end{equation}
where $\bn$ denotes an outward unit normal and square brackets the difference across the interface. 

We follow the standard procedure of linearisation assuming a small, time harmonic, deviation of electron density from equilibrium: $n-n_e \approx \mathrm{Re}\left[e^{-i\omega t}n'\right]$, {and similarly} $\bE \approx \mathrm{Re}\left[e^{-i\omega t}\bE'\right]$, $\bB \approx \mathrm{Re}\left[e^{-i\omega t}\bB'\right]$, and $\bv \approx \mathrm{Re}\left[e^{-i\omega t}\bv'\right]${. Eqs.~\eqref{Gauss}--\eqref{cont NS} governing the metal domain become}
\begin{equation}\label{lin1}
\epsilon_0\bnabla\bcdot \bE'=-en'
\end{equation}
\begin{equation} \label{lin23}
i\omega n' = n_e \bnabla\bcdot \bv', \quad (-i\omega + \gamma)\bv'=-\frac{e}{m}\bE'-\frac{\beta^2}{n_e}\bnabla n'.
\end{equation}
Linearising the interfacial conditions \eqref{bcs} gives 
\begin{equation} \label{bcs lin}
[\bn\bcdot \bE'] =0, \quad [\bn\times \bE']=0, \quad \bn\bcdot\left(\bE'+\frac{m\beta^2}{en_e}\bnabla n'\right)=0. 
\end{equation}
Eliminating $\bv'$ from the formulation by combining \eqref{lin23} gives the scalar equation
\begin{equation}\label{n equation}
\frac{\beta^2}{\omega_p^2}\nabla^2 n'=\left(1-\frac{\omega^2+i\gamma \omega}{\omega_p^2}\right)n',
\end{equation}
where $\omega_p=e\sqrt{n_e/\epsilon_em}$ is the metal's plasma frequency. 
{The subnanometric length scale $\lambda_F=\beta/\omega_p$, roughly $0.2\,$nm for gold, characterises electron-density variations in the metal; we shall refer to it as the nonlocal screening length.} In terms of the ``local'' Drude dielectric function {\cite{Maier:07}},
\begin{equation} \label{Drude}
\epsilon=1-\frac{\omega_p^2}{\omega^2+i\gamma\omega}{,}
\end{equation}
Eq.~\eqref{n equation} equivalently reads as
\begin{equation}\label{n equation 2}
{\lambda^2_F}\nabla^2 n'=\frac{\epsilon}{\epsilon-1}n'.
\end{equation}

In principal, we must supplement Guass's law \eqref{lin1} by the remaining Maxwell equations. At this stage, however, we invoke the quasistatic  approximation \cite{Maier:07}, appropriate for deeply subwavelength plasmonic structures of characteristic size $ \ll c/(\omega \sqrt{|\epsilon|})$, $c$ being the speed of light in vacuum. {The electric near field is then irrotational, $\bnabla\times\bE'\approx0$, 
 allowing us to 
 introduce  an electric potential $\varphi'$ such that $\bE'=-\bnabla\varphi'$. }

\subsection{Surface plasmons and their excitation} \label{ssec:plasmons}
Our formulation governing the near field of the nanometric particle is closed by conditions at (subwavelength) distances large relative to its dimensions{, which depend on the specific scenario under consideration, and in general arise through a matching procedure with the optical far field.} 
For example, to a first approximation, illumination by an electromagnetic plane wave is experienced by a deeply subwavelength particle as an incident uniform time-harmonic electric field. The relevant condition is then $\bE'\sim \unit E_{\infty}$ as $|\boldsymbol{X}|\to\infty$, where $E_{\infty}$ and $\unit$ are the magnitude and polarisation of the incident field, respectively, and $\boldsymbol{X}$ is a position vector relative to some point within the particle. From the properties of Laplace's equation we then have \cite{Jackson:book}
\begin{equation}\label{far exp}
\varphi'\sim -E_{\infty}\,\unit\bcdot\boldsymbol{X} + \frac{\boldsymbol{p}\bcdot \boldsymbol{X}}{4\pi\epsilon_0|\boldsymbol{X}|^3} + \cdots \quad \text{as} \quad |\boldsymbol{X}|\to\infty,
\end{equation}
where the polarisation vector $\boldsymbol{p}$ is an outcome of the near-field problem, from which  quasistatic approximations for the far-field optical cross sections can be derived \cite{Maier:07}. In particular, the absorption cross section in the direction of $\unit$ is 
\begin{equation}\label{Cabs}
\mathcal{C}_{\text{abs}}=\frac{\omega}{c\epsilon_0E_{\infty}}\mathrm{Im}(\boldsymbol{p}\bcdot\unit).
\end{equation}

If, instead of incident radiation, the forcing is in the near field then the vacuum potential $\varphi'$ attenuates at large distances. In \S\ref{sec:sphere} we consider one such example where the forcing is due to a radiating molecule in the vicinity of a nanometallic particle. A radiating molecule is often modelled as an oscillating electric-dipole singularity, with position vector $\boldsymbol{X}_{\boldsymbol{J}}$, hence in this case the Laplace equation governing the vacuum potential $\varphi'$ is replaced by
\begin{equation} \label{dipole dim}
i\omega\epsilon_0\nabla^2\varphi'=-\boldsymbol{J'}\bcdot \bnabla\, \delta_D(\boldsymbol{X}-\boldsymbol{X}_{\boldsymbol{J}}),
\end{equation}
where $\boldsymbol{J}=\mathrm{Re}\left[e^{-i\omega t}\boldsymbol{J}'\right]$ is the  current-density vector and $\delta_D$ denotes the Dirac delta function.  

In the present context, the phenomenon of plasmon resonance crucially relies on the existence, in the absence of external forcing and dissipation ($\gamma=0$), of nontrivial solutions that attenuate at large distances;  this defines a plasmonic eigenvalue problem and we refer to the eigensolutions as the plasmon modes of the metallic nanoparticle. More specifically, for $\omega<\omega_p$, \eqref{n equation} is a modified Helmholtz equation, and eigensolutions are called ``surface plasmons'' since these exhibit an electron-charge distribution confined to a narrow layer adjacent to the surface of the particle. For $\omega>\omega_p$, \eqref{n equation} is a proper Helmholtz equation, and eigensolutions are called ``bulk plasmons''; these exhibit a spatially oscillating electron-density distribution. Modes of the latter type are entirely missed when working in a local formulation where electron-density polarisation is effectively accounted for in terms of a macroscopic dielectric function. 

The physical significance of plasmon eigenmodes stems from the smallness of $\gamma$, for certain plasmonic metals such as gold and silver, relative to typical surface-plasmon eigenfrequencies. Thence, under an external forcing, close to a plasmon eigenfrequency, and having correct symmetries, a damped resonance occurs. That is, to leading order (in $\gamma/\omega$) the near-field distribution mimics that of the corresponding eigenmode, with a large amplitude factor relative to the externally applied field. 

\subsection{Dimensionless formulation} \label{ssec:nondim}
{It is convenient to adopt a dimensionless formulation where lengths are} normalised by a characteristic dimension $a$, and potentials by $aE$, $E$ being a reference field-magnitude value. Specifically, {we define} the dimensionless position vector $\bx=\boldsymbol{X}/a$, potential $\varphi=\varphi'/(aE)$ and the dimensionless {screening length $\delta=\lambda_F/a$}.  From Gauss's law \eqref{lin1} we obtain $n'=O(\epsilon_0E/e\lambda)$, suggesting  a dimensionless charge density $q=-e\lambda n'/(\epsilon_0E)$.  Eqs \eqref{lin1} and \eqref{n equation} governing the metal domain become
\begin{equation} \label{nondim eq}
\delta^2\nabla^2q=\frac{\epsilon}{\epsilon-1}q, \quad \delta\nabla^2\varphi = - q,
\end{equation}
where $\epsilon$ denotes the local Drude function \eqref{Drude}. In vacuum, $q$ is zero and $\varphi$ is governed by the Laplace equation. The interfacial conditions \eqref{bcs lin} become
\begin{equation} \label{nondim bc}
[\varphi]=0,\quad \left[\pd{\varphi}{n}\right]=0, \quad \pd{\varphi}{n}+\delta\pd{q}{n}=0.
\end{equation}

Occasionally it is convenient to use a formulation wherein within the metal we solve for $\chi=\varphi+\delta q$, instead of $\varphi$, and solve Laplace's equation for $\varphi$ in vacuum, and \eqref{nondim eq}, rewritten as
\begin{equation} \label{chi eq}
\delta^2\nabla^2q=\frac{\epsilon}{\epsilon-1}q,\quad \delta\nabla^2\chi=\frac{1}{\epsilon-1}q,
\end{equation}
in the metal domain. The interfacial conditions \eqref{nondim bc} are rewritten as 
\begin{equation} \label{chi bc}
\varphi=\chi-\delta q,\quad \pd{\chi}{n}=0, \quad \pd{\varphi}{n}+\delta\pd{q}{n}=0. 
\end{equation}

\section{Metallic nanosphere}\label{sec:sphere}
\begin{figure}[b]
   \centering
   \includegraphics[scale=0.25]{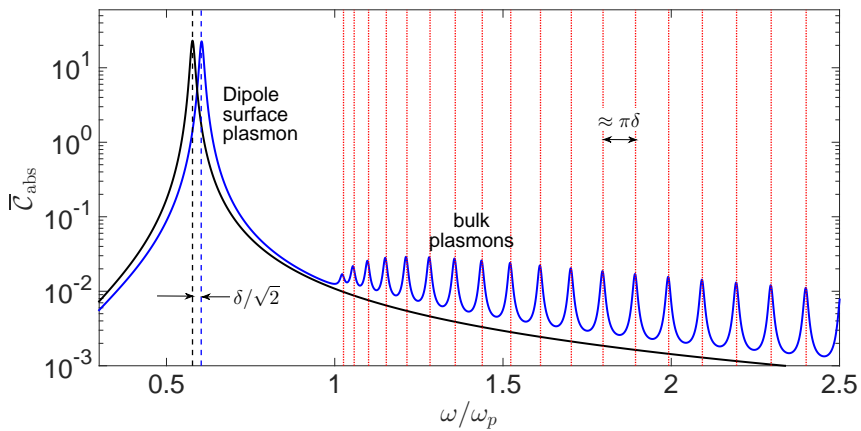} 
   \caption{Normalised absorption cross-section \eqref{normalised abs} for a metal nanosphere (using typical values $a=2$ nm, $\beta=10^6$ m/sec, $\hbar\omega_p=9$ eV and $\hbar\gamma=0.13$ eV \cite{luo:13,Klimov:14}) with $\delta\approx 0.037$
 under plane-wave illumination. Black line ---  local prediction \cite{Maier:07}; blue line --- nonlocal prediction \eqref{mu sphere}; dashed black and blue lines --- local and nonlocal predictions of resonant frequency, respectively [cf.~\eqref{freq shift}]; dotted red lines --- \eqref{high freq} for the high-order bulk-plasmon frequencies.}
   \label{fig:example}
\end{figure}

It is instructive to begin by reviewing the case of a spherical metallic particle, which is amenable to an ``exact'' analytical analysis \cite{Ruppin:73,Klimov:14,Christensen:14,Raza:15}. Emphasis will be placed on approximate simplifications in certain limits, with the intention of motivating the asymptotic approach adopted in later sections.  Considering the scenario of plane-wave illumination as discussed in \S\S\ref{ssec:plasmons}, the dimensionless equations of \S\S\ref{ssec:nondim} are supplemented by the far-field condition
\begin{equation} \label{far}
\varphi \sim -\unit\bcdot\bx + o(1) \quad \text{as} \quad |\bx|\to\infty,
\end{equation}
where the reference length scale $a$ and field magnitude $E$ have been chosen, respectively, as the sphere radius and incident field $E_{\infty}$. Separation of variables provides the deviation of the vacuum potential from the uniform applied field as [cf.~\eqref{far exp}]
\begin{equation} 
\varphi+\unit\bcdot\bx=\mu\frac{\unit\bcdot \bx}{|\bx|^3}, \quad {\rm wherein}\quad 
\mu=\frac{(\epsilon-1)[ik j_1'(ik)-j_1(ik)]}{ik(\epsilon+2)j_1'(ik)+2(\epsilon-1)j_1(ik)}
\label{mu sphere}
\end{equation}
is the dimensionless induced-dipole moment $\boldsymbol{p}\bcdot\unit/(4\pi\epsilon_0a^3E)$. In \eqref{mu sphere}, 
 $j_1$ is the spherical Bessel function of the first kind, and $k(\delta,\epsilon)$ is defined through 
 \begin{equation}
 \delta^2k^2(\epsilon-1)=\epsilon,
 \end{equation}
 with $k>0$ for $\epsilon<0$. 
In Fig.~\ref{fig:example}, for typical parameters used in the literature, we plot the normalised absorption cross-section $\mathcal{\overline C}_{\text{abs}}=\mathcal{C}_{\text{abs}}/[4\pi a^3(\omega_p/c)]$, or 
\begin{equation}\label{normalised abs}
\mathcal{\overline C}_{\text{abs}}=\frac{\omega}{\omega_p}\mathrm{Im}[\mu],
\end{equation}
with $\mu$ given by \eqref{mu sphere}, 
 and by its ``local'' counterpart \cite{Maier:07}
\begin{equation}\label{mu sphere local}
\mu=(\epsilon-1)/(\epsilon+2).
\end{equation}
For $\omega<\omega_p$, plane-wave illumination excites only the fundamental ``dipolar'' surface-plasmon mode of the sphere, which is notably blueshifted from the local-theory prediction represented by the ``Frohlich condition'': $\mathrm{Re}[\epsilon]=-2$. Notably, there are also multiple weak bulk-plasmon modes excited for $\omega>\omega_p$, a feature not captured by a local model. 

From \eqref{mu sphere}, the resonance frequencies of the excited plasmon modes are governed by the transcendental equation
\begin{equation} \label{k eq}
ik(\epsilon+2)j_1'(ik)+2(\epsilon-1)j_1(ik)=0,
\end{equation}
which in general needs to be solved numerically. The typical smallness of the dimensionless Fermi wavelength $\delta$, which is embedded in $k$, suggests  however seeking asymptotic solutions. In fact, the surface-plasmon solution of \eqref{k eq} in the limit $\delta\ll1$ is readily found as $\epsilon\sim -2 + 6\delta\sqrt{{3}/{2}} + O(\delta^2)$, where the leading-order term agrees with the local-theory prediction. Recalling \eqref{Drude}, the corresponding resonance frequency is
\begin{equation} \label{freq shift}
\omega/\omega_p\sim \frac{1}{\sqrt{3}}+\delta\frac{1}{\sqrt{2}}+O(\delta^2).
\end{equation}
The blueshift predicted by \eqref{freq shift} is depicted in Fig.~\ref{fig:example} by dashed vertical lines. As in previous studies \cite{Raza:15}, we derived \eqref{freq shift} by reducing the closed-form solution for a sphere in the limit $\delta\to0$. As we shall see in \S\ref{sec:arbitrary}, it is also possible to alternatively derive \eqref{freq shift} by a direct asymptotic analysis of the nonlocal equations. In fact, one of our main goals is to obtain analogous blueshift formulae for arbitrarily shaped particles for which closed-form solutions do not exist. 

Asymptotic analysis of \eqref{k eq} in the limit $\delta\to0$ also provides the bulk-plasmon frequencies as solutions of the transcendental equation $ik\tan(ik)\sim {2(\epsilon-1)}/({\epsilon+2})$.
Together with \eqref{Drude}, an explicit high-order approximation for the bulk-plasmon frequencies is found as
\begin{equation}\label{high freq}
\omega/\omega_p\sim (1+n^2\pi^2\delta^2)^{1/2}, \quad n \text{ integer } \gg1.
\end{equation}
The predictions of \eqref{high freq} are depicted by the dotted red lines in Fig.~\ref{fig:example}; they accurately pinpoint the resonance peaks, starting from nearly the first bulk resonance. When $n\gg1/\delta$, $\omega/\omega_p\approx n \pi \delta$, and hence the bulk plasmon frequencies become uniformly separated by $\approx \pi\delta$. As far as we are aware, formula \eqref{high freq} for the bulk modes of a sphere is new. 

\begin{figure}[t]
   \centering
   \includegraphics[scale=0.23]{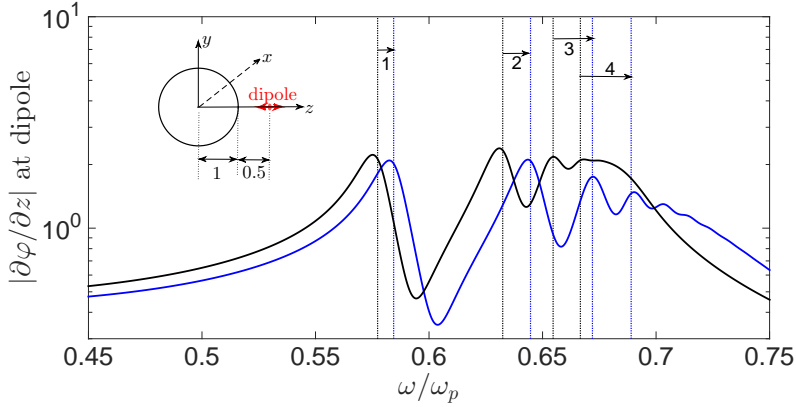} 
   \caption{Multimode excitation of a metal sphere by a nearby radially oriented electric dipole positioned as shown in the inset ($\hbar\omega_p=9$ eV, $\hbar\gamma=0.13$ eV, and $\delta=0.01$). Black and blue lines --- absolute magnitude of the induced radial field at the position of the dipole according to the local and nonlocal models, respectively. Vertical black and blue dotted lines --- first and first two terms of \eqref{sphere corr}.}
   \label{fig:dip}
\end{figure}

For the highly symmetric scenario of plane-wave illumination of a sphere particle, only the dipolar, or Fr\"ohlich, surface-plasmon mode is excited. It is important to emphasise, however, that a nanometallic sphere actually supports an infinite number of  surface plasmons. To demonstrate this we consider a less symmetric forcing in the form of a radiating molecule, modelled as a radially oriented electric dipole placed half a radius away from the particle boundary, with $\boldsymbol{J}'= \mathcal{J}\unit$ in \eqref{dipole dim}. With a reference field value $E=\mathcal{J}/(\epsilon_0\omega_p)$, and defining a dimensionless cartesian system $(x,y,z)$ as shown in Fig.~\ref{fig:dip}, with $\be_z=\unit$, \eqref{dipole dim} governing the vacuum potential becomes
\begin{gather}\label{dipole forcing}
\nabla^2\varphi=i\frac{\omega_p}{\omega}\be_z\bcdot\bnabla \left[\delta_D(x)\delta_D(y)\delta_D(z-l)\right],
\end{gather}
with $\varphi$ attenuating at large distances. While more tedious, separation of variables can here too be applied once the singularity on the right-hand-side of \eqref{dipole forcing} is expanded in spherical harmonics. Of interest is the field induced at the location of the dipole, from which a decay rate can be calculated \cite{Klimov:14,Christensen:14}. In Fig.~\eqref{fig:dip} we show the dimensionless radial field there --- blue line, along with the local-theory prediction --- black line. In contrast to the plane-wave case, several peaks are seen for $\omega<\omega_p$. While the surface-plasmon frequencies can be extracted from the analytic solution (as in e.g.~Ref.~\onlinecite{Raza:15}), we exploit the general theory for arbitrarily shaped  particles to be developed in \S\ref{sec:arbitrary}, which in particular yields the surface-plasmon frequencies of a sphere as 
\begin{equation}\label{sphere corr}
\omega/\omega_p \sim \sqrt{\frac{l}{2l+1}}+\frac{1}{2}\delta\sqrt{l(l+1)} + O(\delta^2), \quad l = 1,2,\ldots
\end{equation}
Eq.~\eqref{sphere corr} agrees with expressions in the literature based on reductions of analytical solutions \cite{Raza:13,Christensen:14,Raza:15}. Note that \eqref{freq shift} is a special case of \eqref{sphere corr} for $l=1$. The applicability of \eqref{sphere corr} is demonstrated in Fig.~\ref{fig:dip}, where the first, and then both (local and then nonlocal) terms of \eqref{sphere corr} are depicted by the vertical dashed black and blue lines, respectively. Notably the effect of nonlocality is stronger for higher-order modes, suggesting the effect is easier to observe under near-field excitation \cite{Christensen:14}. An intuitive explanation is that the relative impact of nonlocality depends on the ratio between the Fermi wavelength $\lambda$ and the characteristic length scale characterising a mode. For low-order modes the latter can be taken to be $a$, here the sphere radius. Higher modes vary more rapidly, and a suitable characteristic scale is $a/l$. This explains the $O(\delta l)$ and $O(\delta n)$ effect in \eqref{sphere corr} and \eqref{high freq}, respectively. Specifically, \eqref{sphere corr} is no longer asymptotic for $l\gg 1/\delta${; such high-order surface plasmons can be excited by electron beams \cite{Raza:15c}, but not with light. The above discussion hints} that nonlocality could play a crucial role when a characterstic scale of the geometry is comparable to $\lambda_F$, a scenario we shall return to in \S\S\ref{ssec:spheroid} and \S\S\ref{ssec:dimer} and consider in detail in \S\ref{sec:near}.

\section{Plasmon blueshift of arbitrarily shaped particles}
\label{sec:arbitrary}
\subsection{{The small-screening-length limit}}
Henceforth our interest is in a direct asymptotic analysis of the nonlocal hydrodynamic equations in the limit $\delta\to0$ with the obvious expectation that, to leading order,  the local electromagnetic approximation is recovered. This requires choosing the length scale $a$ to be the smallest one characterising all relevant modes, and in this section we enforce this by focusing on low-order surface-plasmon modes of metallic nanostructures with a single geometric length scale ${a\gg \lambda_F}$.  An intuitive way to ``derive'' the local approximation from the nonlocal equations is to set $\beta=0$ in \eqref{lin23}, which together with \eqref{lin1} immediately leads to a local relation between electric displacement and field involving the Drude dielectric function \eqref{Drude}, and demanding normal-displacement and tangential-field continuity at the metal-vacuum interface. Such arguments, however, are heuristic, and do not suggest how to systematically go beyond the local approximation.

The results of \S\ref{sec:sphere} imply that nonlocality manifests itself as a surface-plasmon blueshift relative to the resonant frequencies of the local approximation. Motivated by this, 
we now consider the surface-plasmon eigenvalue problem for an arbitrary single-scale particle and ask for what discrete and real frequencies below $\omega_p$ (alternatively, real and negative $\epsilon$ values) does a solution of the nonlocal equations exist with the vacuum potential attenuating at large distances. Following the results of \S\ref{sec:sphere}, we anticipate an eigenvalue expansion in the form
\begin{equation} \label{eps exp}
\epsilon\sim \epsilon\ub{0} + \delta \epsilon\ub{1} + \cdots,
\end{equation}
the corresponding resonance-frequency expansion following from \eqref{Drude} as
\begin{equation} \label{om exp}
\omega/\omega_p\sim \left(1-\epsilon\ub{0}\right)^{-1/2}+\frac{1}{2}\delta\epsilon\ub{1}\left(1-\epsilon\ub{0}\right)^{-3/2}+\cdots.
\end{equation}  
  
Naively, the regular expansion \eqref{eps exp} suggests a comparable expansion of the eigenpotential and charge density. Assuming without loss of generality that the former is $O(1)$, and letting $\bar\varphi$ represent the metal-domain potential, we write 
\begin{equation}\label{bulk exp}
\varphi\sim \varphi\ub{0}(\bx)+\delta\varphi\ub{1}(\bx) + \cdots, \quad \bar\varphi\sim \bar\varphi\ub{0}(\bx)+\delta\bar\varphi\ub{1}(\bx)+\cdots.
\end{equation}
For $\epsilon<0$, it is readily seen from \eqref{nondim eq} that a regular expansion for $q$ would vanish at every algebraic order in $\delta$. On one hand, this implies that $q$ is exponentially small, whereby from \eqref{nondim eq}
\begin{equation}\label{Laplaces}
\nabla^2\varphi\ub{m}(\bx)=0, \quad \nabla^2\bar\varphi\ub{m}(\bx)=0
\end{equation} 
for all orders $m=0,1,2,\ldots$ On the other hand, the interfacial conditions \eqref{nondim bc} demand $\partial q / \partial n =  O(1/\delta)$.
The regular expansion for $\bar\varphi$ in \eqref{bulk exp} must therefore break down at $O(\delta)$ distances from the interface, where \eqref{nondim eq} implies an exponential attenuation of an $q=O(1)$ charge density and consequently a rapid inharmonic variation of $\bar\varphi$. 

The limit $\delta\to0$ is spatially nonuniform and must accordingly be addressed using singular perturbation theory and we employ matched asymptotic expansions to conceptually decompose the metal domain into two regions: (i) An electron-charge boundary layer of width $O(\delta)$ adjacent to the metal-vacuum interface, where $q=O(1)$ and $\bar\varphi$ is inharmonic; and (ii) a bulk-metal domain where $\bx=O(1)$, $q$ is exponentially small, and $\bar\varphi$ is harmonic to all orders. 
The plan is to locally analyse the boundary layer and thereby derive effective boundary conditions connecting the metal and vacuum bulk  domains at successive orders in $\delta$. 
  
\subsection{Boundary-layer analysis} \label{ssec:macro} 
We take an arbitrary smooth particle and grid its boundary by orthogonal unit-metric surface coordinates $(\xi,\eta)$ with unit vectors $(\be_{\xi},\be_{\eta})$.  At a surface point $\mathscr{P}=(\xi,\eta)$, we define a cartesian coordinate system $(x,y,z)$ with origin at $\mathscr{P}$ and unit vectors $(\be_x,\be_y,\be_z)$, where $\be_x,\be_y$ are locally parallel to $\be_{\xi},\be_{\eta}$, and $\be_{z}$ points in the direction of the local outward normal $\bn$. We also define a stretched boundary-layer coordinate $Z=z/\delta$. The boundary-layer fields can be written as
\begin{equation}\label{bl def}
q= Q(\xi,\eta,Z), \quad \chi=\varphi\ub{0}(\xi,\eta)+\delta T(\xi,\eta,Z),
\end{equation}
where bulk fields appearing in boundary-layer equations, e.g.~$\varphi\ub{0}$ in \eqref{bl def},  are understood to correspond to limiting values as the interface is approached, and are accordingly functions of $(\xi,\eta)$ alone. (In the boundary layer it is convenient to work with $\chi$ rather than $\bar\varphi$; note that in the metal bulk these two are the same to exponential order.) A nontrivial subtlety is that $Q$ and $T$ are sought as functions of $Z$ and $(\xi,\eta)$, rather than $Z$ and $(x,y)$. Thus in what follows partial derivatives with respect to $Z$ are with $(\xi,\eta)$, rather than $(x,y)$, held constant. This approach will allow us to conveniently apply boundary conditions at $Z=0$ rather than on a curved surface (for further discussion see e.g.~Ref.~\onlinecite{Cox:97}).

Noting that $\nabla^2\chi=\nabla_s^2\varphi\ub{0}+\delta\nabla^2T$, where $\nabla^2_s$ is the surface Laplacian, \eqref{chi eq} become
\begin{gather}\label{bl eqs}
\delta(\bnabla\bcdot\bn)\pd{Q}{Z}+\pd{^2Q}{Z^2}=\frac{\epsilon}{\epsilon-1}Q+O(Q\delta^2), \\
\delta\left(\nabla_s^2\varphi\ub{0}+\bnabla\bcdot\bn\pd{T}{Z}\right)+\pd{^2T}{Z^2}=\frac{1}{\epsilon-1}Q+O(T\delta^2),
\end{gather}
where the unspecified correction terms correspond to higher order terms in the coordinate transformation. The interfacial conditions \eqref{chi bc} read 
\begin{equation}\label{bl bcs}
\pd{Q}{Z}=-\pd{\varphi}{n}, \quad \pd{T}{Z}=0, \quad \varphi-\varphi\ub{0}=\delta(T-Q) \quad \text{at} \quad Z=0.
\end{equation}
Additional conditions are derived by matching the boundary-layer fields with their metal-bulk counterparts. The matching conditions on $Q$ are attenuation at every algebraic order, as $q$ is exponentially small in the bulk. Those on $T$ are inferred from the Taylor expansion of the bulk potential $\bar\varphi$ in the vicinity of $\mathscr{P}$, rewritten in terms of $Z$. The resulting matching condition reads as 
\begin{multline} \label{matching}
\chi \sim \bar\varphi\ub{0} + \delta\left(Z\pd{\bar\varphi\ub{0}}{n}+\bar\varphi\ub{1}\right)\\ +\delta^2\left[-\tfrac{1}{2}Z^2\left(\bnabla\bcdot\bn\pd{\bar\varphi\ub{0}}{n}+\nabla_s^2\bar\varphi\ub{0}\right)+Z\pd{\bar\varphi\ub{1}}{n}+\bar\varphi\ub{2}\right]+\cdots \quad \text{as} \quad Z\to-\infty.
\end{multline}

We now expand the boundary-layer fields in the form  
\begin{equation}\label{bl exp}
Q\sim Q\ub{0}+\delta Q\ub{1} +\cdots, \quad T\sim T\ub{1}+\delta T\ub{2} + \cdots.
\end{equation}
Substituting \eqref{bl exp} into \eqref{bl eqs} we find at leading order the governing equations 
\begin{equation}\label{Q0T1 eq}
\pd{^2T\ub{1}}{Z^2}=\frac{1}{\epsilon\ub{0}-1}Q\ub{0}, \quad \pd{^2Q\ub{0}}{Z^2}=\frac{\epsilon\ub{0}}{\epsilon\ub{0}-1}Q\ub{0},
\end{equation}
and at first order
\begin{equation}\label{T2 eq}
\pd{^2T\ub{2}}{Z^2}=\frac{1}{\epsilon\ub{0}-1}Q\ub{1}-\frac{\epsilon\ub{1}}{\left(\epsilon\ub{0}-1\right)^2}Q\ub{0}-\nabla_s^2\varphi\ub{0}-\bnabla\bcdot\bn\pd{T\ub{1}}{Z},
\end{equation}
\begin{equation}\label{Q1 eq}
\pd{^2Q\ub{1}}{Z^2}-\frac{\epsilon\ub{0}}{\epsilon\ub{0}-1}Q\ub{1}=-\frac{\epsilon\ub{1}}{\left(\epsilon\ub{0}-1\right)^2}Q\ub{0}-(\bnabla\bcdot\bn)\pd{Q\ub{0}}{Z}.
\end{equation}
Eqs.~\eqref{Q0T1 eq}--\eqref{Q1 eq} are supplemented by the interfacial conditions [cf.~\eqref{bl bcs}] 
\begin{equation}\label{bl bsc A}
\pd{Q\ub{0}}{Z}=-\pd{\varphi\ub{0}}{n}, \quad \pd{Q\ub{1}}{Z}=-\pd{\varphi\ub{1}}{n}, \quad \pd{T\ub{1}}{Z}=\pd{T\ub{2}}{Z}=0, 
\end{equation}
\begin{equation}\label{bl bsc B}
\varphi\ub{1}=T\ub{1}-Q\ub{0} \quad \text{at} \quad Z=0,
\end{equation}
and the matching conditions [cf.~\eqref{matching}] 
{\begin{equation} \label{local pot}
\bar\varphi\ub{0}=\varphi\ub{0} \quad \text{as} \quad Z\to-\infty, 
\end{equation}}
\begin{equation}\label{matching T1}
T\ub{1} \sim Z\pd{\bar\varphi\ub{0}}{n}+\bar\varphi\ub{1} \quad \text{as} \quad Z\to-\infty, 
\end{equation}
\begin{equation}\label{matching T2}
T\ub{2} \sim -\tfrac{1}{2}Z^2\left(\bnabla\bcdot\bn\pd{\bar\varphi\ub{0}}{n}+\nabla_s^2\bar\varphi\ub{0}\right)+Z\pd{\bar\varphi\ub{1}}{n}+\bar\varphi\ub{2}  \quad \text{as} \quad Z\to-\infty.
\end{equation}
Note in particular that \eqref{local pot} serves as our first effective interfacial condition. 
As already noted, matching also entails
\begin{equation}\label{Qto0}
Q\ub{0},Q\ub{1},Q\ub{2},\ldots \to 0 \quad \text{as} \quad Z\to-\infty.
\end{equation}

Solving \eqref{Q0T1 eq} in conjunction with \eqref{bl bsc A} and \eqref{Qto0} gives
\begin{equation}\label{Q0}
Q\ub{0}=-\left(\frac{\epsilon\ub{0}-1}{\epsilon\ub{0}}\right)^{1/2}\pd{\varphi\ub{0}}{n}\exp\left[{\left(\dfrac{\epsilon\ub{0}}{\epsilon\ub{0}-1}\right)^{1/2}Z}\right],
\end{equation}
\begin{equation}\label{T1}
T\ub{1}=-\left(\frac{\epsilon\ub{0}-1}{\epsilon\ub{0}}\right)^{1/2}\frac{1}{\epsilon\ub{0}}\pd{\varphi\ub{0}}{n}\exp\left[{\left(\dfrac{\epsilon\ub{0}}{\epsilon\ub{0}-1}\right)^{1/2}Z}\right]+\frac{1}{\epsilon\ub{0}}\pd{\varphi\ub{0}}{n}Z+\bar\varphi\ub{1}.
\end{equation}
Comparing \eqref{T1} with the matching condition \eqref{matching T1} yields a second effective interfacial condition
\begin{equation}\label{local dis}
\pd{\varphi\ub{0}}{n}=\epsilon\ub{0}\pd{\bar\varphi\ub{0}}{n}.
\end{equation}

As anticipated, the two leading-order effective conditions \eqref{local pot} and \eqref{local dis} are nothing but the familiar ``local'' boundary conditions. To go beyond the local approximation we require two effective interfacial conditions at $O(\delta)$. The first of these, 
\begin{equation}\label{nonlocal pot}
\varphi\ub{1}-\bar\varphi\ub{1}=\left(\frac{\epsilon\ub{0}-1}{\epsilon\ub{0}}\right)^{3/2}\pd{\varphi\ub{0}}{n},
\end{equation}
 follows from potential {continuity \eqref{bl bsc B}}. 
 Towards deriving the second, we integrate \eqref{T2 eq} with respect to $Z$, whereby together with \eqref{Qto0} {and \eqref{T1}} we find
\begin{multline} \label{T2Z}
\pd{T\ub{2}}{Z}\sim -Z\left(\bnabla_s^2\varphi\ub{0}+\frac{1}{\epsilon\ub{0}}\pd{\varphi\ub{0}}{n}\bnabla\bcdot\bn\right) -\frac{1}{\epsilon\ub{0}-1}\int_{-\infty}^0Q\ub{1}\,dZ\\
+\frac{\epsilon\ub{1}}{\left(\epsilon\ub{0}-1\right)^2}\int_{-\infty}^0Q\ub{0}\,dZ
-\bnabla\bcdot\bn\left(\frac{\epsilon\ub{0}-1}{\epsilon\ub{0}}\right)^{1/2}\frac{1}{\epsilon\ub{0}}\pd{\varphi\ub{0}}{n} + o(1) \quad \text{as} \quad Z\to-\infty.
\end{multline}
In \eqref{T2Z}, the integral of $Q\ub{0}$is calculated using \eqref{Q0}, whereas the integral of $Q\ub{1}$ is obtained by integrating \eqref{Q1 eq} with respect to $Z$ together with \eqref{Qto0} and \eqref{bl bsc A}. Eq.~\eqref{T2Z} then gives
\begin{multline}
\pd{T\ub{2}}{Z}\sim -Z\left(\bnabla_s^2\varphi\ub{0}+\frac{1}{\epsilon\ub{0}}\pd{\varphi\ub{0}}{n}\bnabla\bcdot\bn\right) \\+\frac{1}{\epsilon\ub{0}}\pd{\varphi\ub{1}}{n}-\frac{\epsilon\ub{1}}{\epsilon\ub{0}}\pd{\varphi\ub{0}}{n}+ o(1) \quad \text{as} \quad Z\to-\infty.
\end{multline}
Comparing with \eqref{matching T2} we find that the term $\propto Z$ identically matches, whereas matching the $O(1)$ term furnishes the second $O(\delta)$ effective condition as
\begin{equation}\label{nonlocal dis}
\pd{\varphi\ub{1}}{n}-\epsilon\ub{0}\pd{\bar\varphi\ub{1}}{n}=\frac{\epsilon\ub{1}}{\epsilon\ub{0}}\pd{\varphi\ub{0}}{n}. 
\end{equation}

\subsection{Coarse-grained eigenvalue problem}
To summarise the coarse-grained eigenvalue problem{, the bulk potentials $\bar\varphi$ and $\varphi$, expanded as in \eqref{bulk exp}, are governed by Laplace's equation at each order, attenuate at large distances, and satisfy effective interfacial conditions applying at an effective interface, which geometrically coincides with the true vacuum-metal interface.} At leading order these conditions are [cf.~\eqref{local pot} and \eqref{local dis}]
\begin{equation} \label{eff 0}
\varphi\ub{0}=\bar\varphi\ub{0}, \quad \pd{\varphi\ub{0}}{n}=\epsilon\ub{0}\pd{\bar\varphi\ub{0}}{n},
\end{equation}
which, together with attenuation of $\varphi\ub{0}$, define the ``local'' plasmonic eigenvalue problem governing the leading eigenvalues $\epsilon\ub{0}$ and eigenpotentials $(\varphi\ub{0},\bar\varphi\ub{0})$. At $O(\delta)$ the effective conditions are [cf.~\eqref{nonlocal pot} and \eqref{nonlocal dis}]
\begin{equation}\label{eff 1}
\varphi\ub{1}-\bar\varphi\ub{1}=\left(\frac{\epsilon\ub{0}-1}{\epsilon\ub{0}}\right)^{3/2}\pd{\varphi\ub{0}}{n}, \quad \pd{\varphi\ub{1}}{n}-\epsilon\ub{0}\pd{\bar\varphi\ub{1}}{n}={\epsilon\ub{1}}\pd{\bar\varphi\ub{0}}{n}.
\end{equation}
Together with attenuation of $\varphi\ub{1}$, \eqref{eff 1} define a correction problem governing the perturbations $\epsilon\ub{1}$ and $(\varphi\ub{1},\bar\varphi\ub{1})$. Eq.~\eqref{eff 1} shows that, to $O(\delta)$, nonlocality manifests itself macroscopically as an effective potential discontinuity; the local displacement-continuity condition remains valid to this order. 
Determining the eigenvalue correction $\epsilon\ub{1}$ does not require a detailed solution of the  $O(\delta)$ problem. Rather, as we show next, $\epsilon\ub{1}$ can be obtained directly from knowledge of the  ``local'' eigenvalues and eigenpotentials. In essence, our coarse-graining procedure has regularised the otherwise singular small-$\delta$ limit, allowing us to apply standard ideas of regular perturbation theory.

\subsection{Nonlocal perturbation from simple and degenerate eigenvalues}
We first take $\epsilon\ub{0}$ as a simple eigenvalue linked with one distinct eigenpotential pair $(\varphi\ub{0},\bar\varphi\ub{0})$. Since attenuation at large distances and Laplace's equation hold at each algebraic order, the correction problem [cf.~\eqref{eff 1}] is a forced version of the leading-order local problem [cf.~\eqref{eff 0}]. The former can therefore posses a solution only under special circumstances. Indeed, applying Green's second identity to the pairs $(\varphi\ub{0},\varphi\ub{1})$ and $(\bar\varphi\ub{0},\bar\varphi\ub{1})$, and using attenuation and \eqref{eff 0} and \eqref{eff 1}, we find the solvability condition
\begin{equation} \label{eps 1}
\epsilon\ub{1}=\left(\epsilon\ub{0}\right)^2\left(\frac{\epsilon\ub{0}-1}{\epsilon\ub{0}}\right)^{3/2}\dfrac{\oint\left[\pd{\bar\varphi\ub{0}}{n}\right]^2\,dA}{\oint\bar\varphi\ub{0}\pd{\bar\varphi\ub{0}}{n}\,dA},
\end{equation}
where integrals are taken over the effective metal-vacuum interface. By applying Green's first identity to the denominator it is readily seen that $\epsilon\ub{1}>0$, i.e.~hydrodynamic nonlocality results in a surface-plasmon blueshift for arbitrary shaped particles. 

There are many cases, particularly for highly symmetric configurations, where more than one independent plasmon mode is supported at a single eigenfrequency. The preceding argument is readily generalised to such cases  where $\epsilon\ub{0}$ is $N$-times degenerate, with distinct eigenpotential pairs $(\varphi\ub{0}_n$,$\bar\varphi\ub{0}_n)$, where $n=1,2,..,N$.  To allow for a general ``local'' state, we recast \eqref{bulk exp} as
\begin{equation}
\varphi\sim \sum_{n=1}^N\alpha_n\varphi_n\ub{0}(\bx)+\delta\varphi\ub{1}(\bx) + \cdots, \quad \bar\varphi\sim \sum_{n=1}^N\alpha_n\bar\varphi_n\ub{0}(\bx)+\delta\bar\varphi\ub{1}(\bx) + \cdots, 
\end{equation}
where $\{\alpha_n\}$ is a set of real numbers to be determined together with $\epsilon\ub{1}$. 
Applying Green's second identity to the $2N$ pairs $(\varphi\ub{1},\varphi_m\ub{0})$ and $(\bar\varphi\ub{1},\bar\varphi_m\ub{0})$, where $m=1\ldots N$,
 and using attenuation, 
\begin{equation} \label{new ev}
\left[\epsilon\ub{0}\right]^2\left(\frac{\epsilon\ub{0}-1}{\epsilon\ub{0}}\right)^{3/2}\sum_{n=1}^N\alpha_n\oint\pd{\bar\varphi_n\ub{0}}{n}\pd{\bar\varphi_m\ub{0}}{n}\,dA=\epsilon\ub{1}\sum_{n=1}^{N}\alpha_n\oint\bar\varphi_m\ub{0}\pd{\bar\varphi_n\ub{0}}{n}\,dA.
\end{equation}
{Eq.~\eqref{new ev} is a} matrix problem for the eigenvalues $\epsilon\ub{1}$ and the eigenvectors $\{\alpha_n\}$.  
We shall next apply \eqref{eps 1} and \eqref{new ev} to geometries for which the leading ``local'' eigenpotentials and eigenvalues have been obtained in the literature using separation of variables. 

\subsection{The sphere and circular cylinder}
\label{ssec:sphereagain}

Returning to the sphere example of \S\ref{sec:sphere}, 
in the local approximation the plasmonic eigenvalues and eigenpotentials of a sphere are particularly simple and are well known to be
\begin{equation} \label{sphere local}
\epsilon\ub{0}_l = -\frac{l+1}{l}, \quad l=1,2,\ldots, 
\end{equation}
and
\begin{equation}\label{sphere pot}
\bar\varphi\ub{0}_{lm}=r^lY_{lm}(\theta,\phi), \quad m=-l,\ldots, l,
\end{equation}
where $(r,\theta,\phi)$ are arbitrarily oriented spherical coordinates with origin at the sphere centre and $Y_{lm}$ are spherical harmonics; the external potentials are  not required. Since $\epsilon\ub{0}_l$ is $(2l+1)$ degenerate, we must in principal employ \eqref{new ev}, but the integrals are zero for $n\neq m$ from the orthogonality properties of $\bar\varphi\ub{0}_{lm}$. The linear system \eqref{new ev} therefore reduces to \eqref{eps 1}, yielding 
\begin{equation} \label{sphere correction}
\epsilon_l\ub{1} = l^{-1}(l+1)^{1/2}(2l+1)^{3/2},
\end{equation}
with the $(2l+1)$ degeneracy surviving to $O(\delta)$. Note that rewriting \eqref{sphere local} and \eqref{sphere correction} in terms of frequency using \eqref{om exp} yields \eqref{sphere corr}. 

\begin{figure}[t]
   \centering
   \includegraphics[scale=0.4]{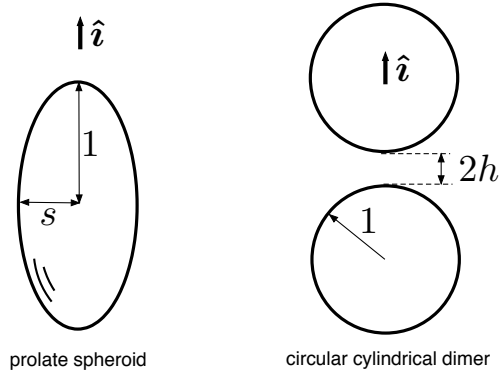} 
   \caption{Dimensionless schematics of nanometallic configurations considered in \S\ref{sec:arbitrary} and \S\ref{sec:near}.}
   \label{fig:shapes}
\end{figure}

Consider next the case of a circular cylinder of dimensionless radius $1$. Local theory yields just one eigenvalue, $\epsilon\ub{0}=-1$, which is infinitely degenerate with internal eigenpotentials 
\begin{equation}\label{cyl pot}
\bar\varphi\ub{0}_{m0}=r^m\cos (m\phi), \quad \bar\varphi\ub{0}_{m1}=r^m\sin (m\phi), \quad m=1,2,\ldots
\end{equation}
Eqs.~\eqref{new ev} furnish the system of equations
\begin{equation}
2^{3/2}m\alpha_{mp}=\epsilon\ub{1}\alpha_{mp}, \quad p=0,1, \quad m=1,2,\ldots, 
\end{equation}
which has an infinite number of eigenvalue solutions
\begin{equation} \label{eps 1 cyl}
\epsilon\ub{1}_m=2^{3/2}m,\quad m=1,2,\ldots,
\end{equation}
where for each $m$ the corresponding eigenvector satisfies $\alpha_{np}=0$ for $n\ne m$. Interestingly, at $O(\delta)$ the unique local eigenvalue $\epsilon\ub{0}$ splits into an infinite set of eigenvalues, each remaining only doubly degenerate. The inapplicability of the regular perturbation scheme for large $m$ does not affect the perturbation for small and moderate $m$ owing to the decoupling in \eqref{new ev} for different $m$ (but not $p$) values. Justified suspicion arising from use of \eqref{new ev} for infinite degeneracy is mitigated by a comparison with an exact separable solution of the nonlocal model in polar coordinates.

\subsection{Prolate spheroid}
\label{ssec:spheroid}
\begin{figure}[t]
   \centering
   \includegraphics[scale=0.25]{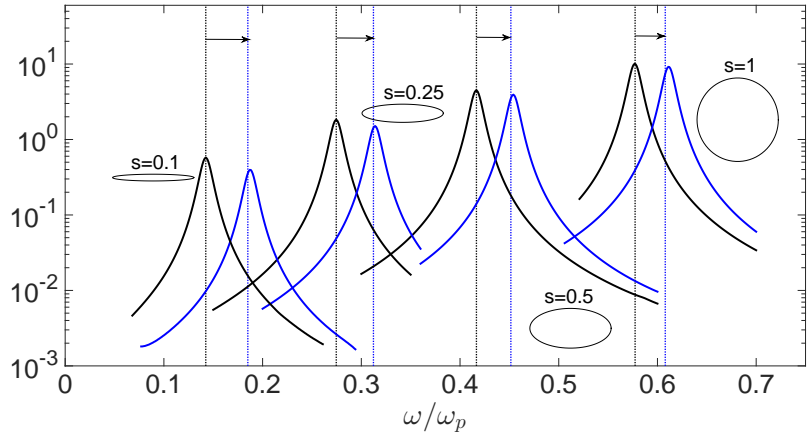} 
   \caption{Dimensionless near field $|\mathrm{Im}[\unit\bcdot\bnabla\varphi]|$ external to metallic prolate spheroid ($\gamma/\omega_p=0.014$, $\delta=0.04$) subjected to a plane wave polarised along $\unit$, for several values of the slenderness parameter $s$, see Fig.~\ref{fig:shapes} (the field is probed on the revolution axis at unit distance from the spheroid tip). Black and Blue lines are respectively the local and nonlocal responses from a quasistatic simulation. 
 The vertical dotted black and blue lines respectively depict the dipolar resonant frequency predicted from local theory [cf.~\eqref{eps 0 s}] and our perturbative nonlocal theory [cf.~\eqref{eps 1 s}].}
   \label{fig:spheroid}
\end{figure}

As a nontrivial example, consider a prolate spheroid with dimensionless semi-axes $1$ and $s<1$ (see Fig.~\ref{fig:shapes}).  In the local approximation, separation of variables in prolate-spheroidal coordinates \cite{Klimov:14} provides the eigenvalues as 
\begin{equation} \label{eps spheroid}
\epsilon\ub{0}_{nm}=\left.\frac{{Q_n^m}'(x)P_n^m(x)}{{P_n^m}'(x)Q_n^m(x)}\right|_{x=1/\sqrt{1-s^2}},  \quad n=1,2,\ldots,\quad m=0,1,\ldots,n,
\end{equation}
where $P_n^m$ and $Q_n^m$ are associated Legendre functions of the first and second kind, with a branch cut from $-\infty$ to $1$ along the real axis. Eigenvalues with $m=0$ correspond to axisymmetric modes, and are simple; eigenvalues with $m\ne0$ are non-axisymmetric and are twice degenerate in accordance with the rotational symmetry of the spheroid. In particular, the eigenvalue $\epsilon\ub{0}_{10}$ can be expressed explicitly as \footnote{Note that for $x>1$:  $P_1^0(x)=x, \,  Q_1^0(x)=\frac{1}{2}x\ln\frac{x+1}{x-1}-1$.}
\begin{equation} \label{eps 0 s}
\epsilon\ub{0}_{10}=\frac{\ln\frac{1+\sqrt{1-s^2}}{1-\sqrt{1-s^2}}-2s^{-2}\sqrt{1-s^2}}{\ln\frac{1+\sqrt{1-s^2}}{1-\sqrt{1-s^2}}-2\sqrt{1-s^2}}.
\end{equation}
The associated surface-plasmon eigenfield is uniform within the metal \cite{Klimov:14}, that is $\bar\varphi\ub{0}_{10}=\unit\bcdot\bx $, where $\unit$ is a unit vector parallel to the spheroid's major axis. Invoking prolate-spheroidal coordinates to evaluate the integrals in \eqref{eps 1}, the nonlocal eigenvalue perturbation is found as
\begin{equation}\label{eps 1 s}
\epsilon\ub{1}_{10}=\frac{3}{2}\left[\epsilon\ub{0}_{10}\right]^2\left(\frac{\epsilon\ub{0}_{10}-1}{\epsilon\ub{0}_{10}}\right)^{3/2}\frac{s}{(1-s^2)^{3/2}}\left[\arctan\frac{\sqrt{1-s^2}}{s}-s\sqrt{1-s^2}\right].
\end{equation} 
Eq.~\eqref{eps 1 s}, together with \eqref{om exp} and \eqref{eps 1 s}, provides the blueshift of the fundamental axisymmetric, Fr\"ohlich-type, surface-plasmon mode of a prolate spheroid. The latter mode is of special importance in sensing applications \cite{Link:99}, as it allows a tuneable plasmon frequency over a broad band and high field enhancements; it is also the only mode which can be effectively excited by light polarised parallel to the axis. The correction \eqref{eps 1 s} may be of particular importance for accurately utilising elongated particles as ``plasmon rulers''. To demonstrate the applicability of \eqref{eps 1 s} we have numerically solved, using the commercial finite element package Comsol Multiphysics\textsuperscript{\textregistered}, the problem of a subwavelength (quasistatic) prolate spheroid illuminated polarised parallel to $\unit$. In Fig.~\ref{fig:spheroid} we plot, for $s=1,0.5,0.25$ and $0.1$, the imaginary part of the axial field included on the symmetry axis at unity distance away from the spheroid tip. The black and blue lines respectively depict the local and nonlocal calculations. The vertical dashed black and blue lines respectively mark the eigenfrequency predictions associated with \eqref{eps spheroid} and \eqref{eps 1}.

Fig.~\ref{fig:spheroid} demonstrates the familiar geometrical redshift as $s\to0$ in the local plasmon frequency. Indeed, analysis of the local approximation \eqref{eps 0 s} shows that 
\begin{equation}\label{spheroid s}
\epsilon\ub{0} \sim  -\frac{1}{\left[\ln({2}/{s})-1\right]s^2}  \quad \text{as} \quad s\to0,
\end{equation}
i.e.~in the local approximation the plasmon frequency redshifts rapidly and without bound as $s\to0$. Adding to \eqref{spheroid s} the small-$s$ asymptotics of the nonlocal perturbation \eqref{eps 1 s}, we find, formally,
\begin{equation}\label{dubious s}
\epsilon \sim \frac{1}{\left[\ln({2}/{s})-1\right]s^2}\left[-1+\frac{3\pi}{4}\frac{\delta/s}{\ln(2/s)-1}+\cdots\right], \quad \delta\ll s\ll1.
\end{equation}
The asymptotic hierarchy in \eqref{dubious s} is clearly invalidated  when $s=O(\delta)$. Indeed, in this extreme the transverse metal dimension is $O(\lambda_F)$ and we expect nonlocality to be appreciable. We next turn to an example where nonlocality becomes dominant even though the metal dimensions are strictly $\gg\delta$, owing to a geometric {electric-field amplification}. 

\subsection{Circular cylindrical dimer}
\label{ssec:dimer}
As a final exemplar of the perturbation formula \eqref{eps 1} consider a pair of identical metallic cylinders of dimensionless radius $1$, separated by a gap of dimensionless width $2h$, see Fig.~\ref{fig:shapes}. 
The ``local'' eigenvalues and eigenpotentials are derived in Ref.~\onlinecite{Klimov:14} by  separation of variables in bipolar coordinates; the eigenpotentials are characterised as being either symmetric or antisymmetric about the plane bisecting the gap. We focus on  the antisymmetric modes, known also as the longitudinal or ``bonding'' modes, that 
are particularly important in plasmonic applications owing to their tuneability and high field confinement. The relevant eigenvalues are
\begin{equation}\label{dimer exact}
\epsilon\ub{0}_n=-\coth\left[(n+1)\cosh^{-1}(1+h)\right], \quad n=0,1,2,\ldots
\end{equation}
each of which is doubly degenerate, with associated bonding modes symmetric and anti-symmetric about an axis coinciding with the line of centres. Conveniently, however, the eigenpotentials as given in Ref.~\onlinecite{Klimov:14} are orthogonal over the metal-vacuum interface, whereby the system of equations \eqref{new ev} again degenerates to \eqref{eps 1}. Evaluating the integrals in bipolar coordinates, the latter yields
\begin{equation} \label{eps 1 d}
\epsilon\ub{1}_n = -(n+1)\epsilon\ub{0}_0\left[\epsilon\ub{0}_n\right]^2\left[\frac{\epsilon\ub{0}_n-1}{\epsilon\ub{0}_n}\right]^{3/2}.
\end{equation}
This result is demonstrated in Fig.~\ref{fig:nonlocal_freq}, which will be discussed in \S\ref{sec:macro}. 

The bonding modes of a cylindrical dimer redshift as $h\to0$ analogously to the longitudinal Fr\"ohlich mode of a prolate spheroid as $s\to0$ [cf.~\eqref{spheroid s}]. Inspection of \eqref{dimer exact} shows that \cite{Vorobev:10}
\begin{equation} \label{dimer h}
\epsilon\ub{0}_n\sim -\frac{1}{n+1}\frac{1}{\sqrt{2h}}+O(\sqrt{h}), \quad \text{as} \quad h\to0.
\end{equation}
Adding to \eqref{dimer h} the small-$h$ asymptotics of \eqref{eps 1 d} we find, formally, 
\begin{equation}\label{dimer break}
\epsilon \sim -\frac{1}{n+1}\frac{1}{\sqrt{2h}}+\frac{\delta}{n+1}\frac{1}{(2h)^{3/2}}, \quad \delta\ll h \ll1,
\end{equation}
which is no longer asymptotic when $h=O(\delta)$. The  breakdown {here} of the regular eigenvalue expansion \eqref{eps exp} is, unlike {in the spheroid case}, linked to a nonmetallic rather than a metallic dimension becoming comparable {to the screening length}.

\section{Nearly touching circular cylinders}
\label{sec:near}
In this section we revisit the plasmonic eigenvalue problem for a circular cylindrical dimer (see \S\S\ref{ssec:dimer}), this time carrying out an \emph{ab initio} asymptotic analysis in the prevailing case where both $h$ and $\delta$ are small. As in \S\S\ref{ssec:dimer}, we shall focus on bonding eigenpotentials, which are antisymmetric with respect to the plane bisecting the gap. For $h\gg \delta$, we expect the small-perturbation scheme of \S\S\ref{ssec:dimer} to hold; thus, in this case \eqref{eps 1 d} is valid to leading order, with nonlocality playing only a relatively minor role captured by the second term in \eqref{dimer break}. In contrast, in the strongly nonlocal case, $h= O(\delta)$, we expect nonlocality to affect the plasmon eigenfrequencies at leading order. Before tackling the strongly nonlocal limit, we shall first in \S\S\ref{ssec:dimerlocal} derive the local near-contact asymptotics \eqref{eps 1 d}, adopting the singular perturbations approach demonstrated in Ref.~\onlinecite{Schnitzer:15:dimer} for a sphere dimer (in the local approximation). We shall then find in \S\S\ref{ssec:dimernonlocal} that this approach extends in a simple way to the strongly nonlocal case, where an exact analytical approach is intractable. 

\subsection{Weak nonlocality: $h\gg\delta$}\label{ssec:dimerlocal}
\subsubsection{Local eigenvalue problem}
To derive the leading order surface-plasmon eigenvalues and eigenpotentials in the near-contact limit $h\ll1$, with $\delta\ll h$, we disregard nonlocality and seek nontrivial eigenpotentials $\varphi$ and $\bar\varphi$ satisfying Laplace equations in the vacuum and metal domains, respectively, along with %
\begin{equation} \label{local bcs}
\varphi=\bar\varphi, \quad \pd{\varphi}{n}=\epsilon\pd{\bar\varphi}{n}
\end{equation}
as the local interfacial conditions 
and attenuation of $\varphi\to0$ at large distances. We introduce cartesian coordinates    $(x,y)$, with unit vectors ($\be_x$,$\be_y$), origin at the gap centre, and $\be_y$ directed parallel to the line of particle centres (see Fig.~\ref{fig:2dschematic}). Our interest lies in bonding modes satisfying
\begin{equation}\label{symm}
\varphi(x,y)=-\varphi(x,-y),
\end{equation}
and similarly for $\bar\varphi$.  Note that (\ref{symm}) allows us to consider only the half plane $y>0$, say, by prescribing the condition
\begin{equation}\label{as 2d}
\varphi(x,0)=0.
\end{equation}

The scaling of the $\epsilon$ eigenvalues as $h\to0$ follows from arguments similar to those given in Refs.~\cite{Schnitzer:15:dimer} and \cite{Schnitzer:16}. 
{We define $(x,y)=O(h^{1/2},h)$ as the gap region, where given the locally parabolic cylinder boundaries the gap separation remains $O(h)$. Assuming $\varphi=O(1)$ there, \eqref{as 2d} implies an $O(1/h)$ transverse gap field. Also, since  
 for $y=O(h)$ and  $|x|\gg h^{1/2}$, \eqref{as 2d} implies $\varphi=O(h)$, and hence for fixed $y$ the gap potential attenuates over $x=O\left(h^{1/2}\right)$.} Considering next the metal side of the gap boundary, continuity of potential implies $\bar\varphi$ is $O(1)$  varying rapidly over $O\left(h^{1/2}\right)$ distances along the boundary. Owing to the apparent unboundedness of the metal domain on such small scales, and the symmetry of Laplace's equation, we expect equally rapid transverse variations in potential, i.e.~the transverse pole field is $O(1/h^{1/2})$; we refer to the $O\left(h^{1/2}\right)\times O\left(h^{1/2}\right)$ metal regions as the poles. Substituting the transverse-field scalings in the gap and pole domains into the electric displacement condition at the vacuum-metal interface suggests
\begin{equation} \label{eps scaling}
\epsilon\sim -  \zeta \,h^{-1/2},
\end{equation}
where $\zeta=O(1)$ is a prefactor to be determined. The asymptotic structure of the sought eigenpotentials is closed by noting that, since $|\epsilon|\gg1$, on $O(1)$ length scales, away from the gap, the metal cylinders must to leading order be at fixed and opposite potentials, say $\pm\mathcal{V}$. 

\subsubsection{Asymptotic analysis}
\begin{figure}[t]
   \centering
   \includegraphics[scale=0.6]{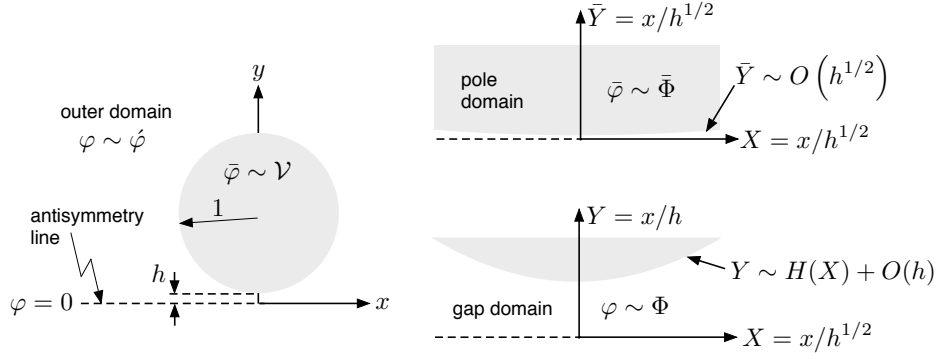} 
   \caption{Near-contact asymptotic structure of bonding modes of a circular cylindrical dimer in the local approximation.}
   \label{fig:2dschematic}
\end{figure}

More formally, the gap domain 
 is defined to be where the stretched coordinates $X=x/h^{1/2}$ and $Y=y/h$ are $O(1)$ (see Fig.~\ref{fig:2dschematic}), the cylindrical 
 boundaries reading as $Y\sim \pm H(X) + O(h)$, where $H(X)=1+\frac{1}{2}X^2$. 
Expanding the gap potential as $\varphi\sim \Phi(X,Y) + O(h^{1/2})$, Laplace's equation at leading order degenerates to 
$\partial^2\Phi/\partial Y^2=0$, 
which together with \eqref{as 2d} shows that
\begin{equation} \label{Phi}
\Phi=-E(X)Y.
\end{equation}
The gap field is dominantly transverse 
 with the field distribution $E(X)$ yet to be determined. 

Consider next the pole domain within the metal cylinder (in $y>0$, say), where the alternative set of stretched coordinates, ${X}$ and $\bar{Y}=y/h^{1/2}$, are $O(1)$. With this stretching, the cylinder boundary appears flat, $\bar{Y}\sim O(h^{1/2})$.
Expanding the potential there as $\varphi\sim \bar{\Phi}(X,\bar{Y}) + O(h^{1/2})$, Laplace's equations gives
\begin{equation} \label{cyl pole eq}
\pd{^2\bar\Phi}{{X}^2}+\pd{^2\bar\Phi}{\bar{Y}^2}=0,
\end{equation}
whereas the interfacial conditions \eqref{local bcs}, in conjunction with \eqref{Phi}, read 
\begin{equation} \label{cyl connection}
\bar{\Phi}=-E(X)H(X), \quad \zeta\pd{\Bar \Phi}{\bar Y}=E(X),\quad \text{at} \quad \bar{Y}=0.
\end{equation}
The latter can be combined to form the mixed-type boundary condition
\begin{equation} \label{cyl mixed}
\bar\Phi+\zeta H(X)\pd{\bar\Phi}{\bar{Y}}=0 \quad \text{at} \quad \bar{Y}=0.
\end{equation}
Eqs.~\eqref{cyl pole eq} and \eqref{cyl mixed}, together with the matching condition
$\bar\Phi\to\mathcal{V}$ as $X^2+\bar{Y}^2\to\infty$,  
define an effective eigenvalue problem in the half plane $\bar{Y}>0$, which we shall now solve.

It is convenient to work with the potential deviation $\psi=\bar\Phi-\mathcal{V}$, and its Fourier transform,
\begin{equation}
\hat{\psi}(k,\bar{Y})=\frac{1}{\sqrt{2\pi}}\int_{-\infty}^{\infty}\psi\left(X,\bar{Y}\right)e^{-ikX}\,dX.
\end{equation}
Eqs.~\eqref{cyl pole eq} and the matching condition at infinity together imply 
\begin{equation}
\hat{\psi}=M(k)e^{-|k|\bar{Y}}.
\end{equation}
Fourier transforming the mixed-type boundary condition \eqref{cyl mixed}, we find that the distribution $M(k)$ is governed by the differential equation
\begin{equation} \label{M eq}
(1-\zeta |k|)M(k)+\frac{\zeta}{2}\frac{d^2}{dk^2}\left[|k|M(k)\right]=-\sqrt{2\pi}\mathcal{V}\delta_D(k),
\end{equation}
$\delta_D(k)$ denoting the Dirac-delta function. 

It is convenient at this point to consider separately eigenpotentials even and odd in $x$, for which $M(k)$ is, respectively, even and odd in $k$. For eigenpotentials even in $x$ we  write \eqref{M eq} as 
\begin{equation}
\frac{d^2(kM)}{dk^2}+2\left(\frac{1}{\zeta}-k\right)M=0, \quad k>0,
\end{equation}
together with the condition
\begin{equation} \label{delta cond}
M(0)=-\sqrt{2\pi}\mathcal{V}/\zeta, 
\end{equation}
which is obtained by balancing the singularity at $k=0$. A large-$k$ analysis suggests the substitution $M(k)=e^{-\sqrt{2}|k|}N(2\sqrt{2}|k|)$, whereby
\begin{equation}\label{cyl T}
p\frac{d^2N}{dp^2}+(2-p)\frac{dN}{dp}-\left(1-\frac{1}{\sqrt{2}\zeta}\right)N=0, \quad p=2\sqrt{2}k>0.
\end{equation}
Eq.~\eqref{cyl T} is identified as the associated Laguerre differential equation. Nonsingular solutions consistent with \eqref{delta cond} and having algebraic growth as $p\to\infty$ exist only when the bracketed term in \eqref{cyl T} is a nonpositive integer, say $n$, the solution then proportional to $L^1_n(p)$, the $n$th associated Laguerre polynomial of order one \cite{Abramowitz:book}. Thence the leading-order eigenvalues are
\begin{equation} \label{cyl alpha}
\zeta_n=\frac{1}{\sqrt{2}(1+n)}, \quad n=0,1,2,\ldots
\end{equation}
Noting that $L^1_n(0)=n+1$, the corresponding eigenfunctions $M=M^e_n(k)$ satisfying \eqref{delta cond} are
\begin{equation}\label{Me}
M^e_n(k) = -2\sqrt{\pi}\mathcal{V}e^{-\sqrt{2}|k|}L^1_n(2\sqrt{2}|k|);
\end{equation}
the outer voltage $\mathcal{V}$ is an arbitrary multiplicative factor determining the magnitude of the mode. 

As prompted, there are also eigensolutions of \eqref{M eq} that are odd in $k$, with $\mathcal{V}=0$. If $M$ is to remain finite as $|k|\to0$ [this is justified in the discussion below \eqref{outer pot}], then again solutions with appropriate behaviour as $|k|\to\infty$ are only possible for $\zeta$ eigenvalues given by \eqref{cyl alpha}, which are accordingly degenerate as already mentioned in \S\S\ref{ssec:dimer}. The odd transformed solutions are readily seen to be the discontinuous variants of \eqref{Me}, 
\begin{equation}\label{Mo}
M^o_n(k) = -2\sqrt{\pi}\mathcal{D}\mathrm{sg}(k)e^{-\sqrt{2}|k|}L^1_n(2\sqrt{2}|k|),
\end{equation}
where, with $\mathcal{V}=0$, we take $\mathcal{D}$ as an arbitrary multiplicative factor. 

\begin{figure}[t]%
       \centering
         \includegraphics[scale=0.45]{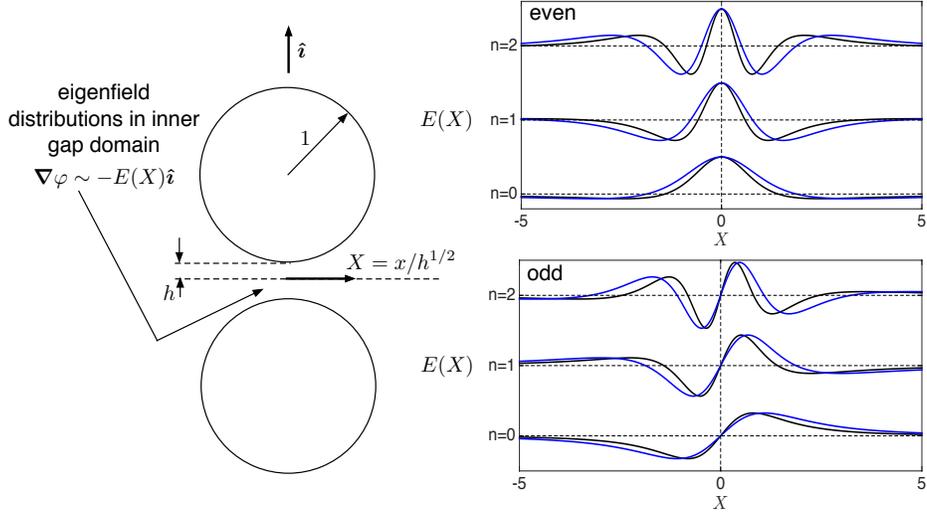} 
       \caption{Gap-field distributions $E(X)$ [cf.~\eqref{Phi}] corresponding to the first three resonance frequencies. Each eigenfrequency is doubly degenerate with eigenfields even and odd in $X$. The black lines are the ``local'' distributions \eqref{E even} and \eqref{E odd}. The blue lines depict the ``nonlocal'' renormalised distributions [cf.~\eqref{trans}] at the blueshifted frequencies \eqref{nonlocal renorm}, for $\tau=\delta/h=0.75$. }
              \label{fig:modeseo}
     \end{figure}

The eigenpotentials may be readily calculated by explicit Fourier inversion. Of most interest perhaps is the $x$-distribution of the gap field [cf.~\eqref{Phi}],
\begin{equation}
E^{e/o}_n(X)=-\frac{\zeta_n}{\sqrt{2\pi}}\int_{-\infty}^{\infty}|k|M^{e/o}_n(k)e^{ikX}\,dk.
\end{equation}
Noting that $L^1_0(p)=1,L^1_1(p)=2-p$, and $L^1_2(p)=3-3p+p^2/2$, the first three even gap-field distributions are
\begin{multline}\label{E even}
E^e_0(X)/\mathcal{V}=\frac{2(2-X^2)}{(2+X^2)^2}, \quad E^e_1(X)/\mathcal{V}=-\frac{2(4-12X^2+X^4)}{(2+X^2)^3},\\ E^e_2(X)/\mathcal{V}=-\frac{2(-8+60X^2-30X^4+X^6)}{(2+X^2)^4}, \ldots,
\end{multline}
while the first three odd ones are
\begin{multline}\label{E odd}
E^o_0(X)/(i\mathcal{D})=\frac{4\sqrt{2}X}{(2+X^2)^2}, \quad E^o_1(X)/(i\mathcal{D})=\frac{8\sqrt{2}X(X^2-2)}{(2+X^2)^3}, \\ E^o_2(X)/(i\mathcal{D})=\frac{4\sqrt{2}X(12-20X^2+3X^4)}{(2+X^2)^4}, \ldots
\end{multline}
The above distributions are depicted by the black lines in Fig.~\ref{fig:modeseo}. We note that the even modes are somewhat less strongly confined in the gap than the odd ones. In fact, in the three dimensional case of a sphere dimer, the attenuation of the axisymmetric modes (analogous to the even modes here) is so slow that the eigenfrequencies cannot be determined by analysing the gap and pole domains separately from the particle-scale field distribution \cite{Schnitzer:15:dimer}. 

To complete the description of the eigenpotentials we consider the ``outer'' vacuum domain, at $O(1)$ distances away from the gap where to leading order the cylinders appear to be touching. For eigenpotentials even in $x$, alluding to the analysis of nearly touching perfectly conducting cylinders (see appendix of Ref.~\onlinecite{Jeffrey:78}), it is readily verified that the outer vacuum potential is
\begin{equation}\label{outer pot}
\varphi\sim\acute\varphi + \cdots, \quad \acute\varphi = 2\mathcal{V}\frac{y}{x^2+y^2}. 
\end{equation}
Indeed, this distribution satisfies Laplace's equation, is even in $x$ and odd in $y$, attenuates at large distances, and takes the value $\mathcal{V}$ at the approximate boundary of the upper cylinder. Importantly, \eqref{outer pot} asymptotically matches with the gap region. Thus, writing $\acute\varphi$ in terms of the inner gap coordinates and expanding for small $h$ gives $\sim 2\mathcal{V}Y/X^2$, which in turn implies $E(X)\sim -2\mathcal{V}/X^2$; it is easily seen that the latter is the large $|X|$ behaviour of the gap distributions \eqref{E even} for all $n$. Consider next the eigenpotentials odd in $x$, for which $\mathcal{V}=0$. Eqs.~\eqref{E odd} imply that as $X\to\infty$, $E(X)\sim O(1/X^3)$ and hence from \eqref{Phi} the potential is $O(Y/X^3)$. Matching in turn suggests the outer potential is $O(h^{1/2})$ small. The seemingly trivial $O(1)$ outer problem involving Laplace's equation, a homogeneous Dirichlet condition on the cylinder boundaries, and attenuation at large distances, actually possesses solutions which are singular at the origin \cite{Jeffrey:78}; these, however, are too singular to match any allowable inner solution;
 this in turn justifies \textit{a posteriori} our assumption that $M(k)$ is finite at the origin.

\subsection{Strong nonlocality: $h = O(\delta)$} \label{ssec:dimernonlocal}
With the near-contact asymptotics in the local approximation determined, we proceed to consider the strongly nonlocal limit, where the gap width is small compared with the cylinder radius and comparable to the {nonlocal screening length}, i.e.~$1\gg h=O(\delta)$. From the coarse-graining procedure of \S\S\ref{ssec:macro} we expect that away from the gap, where the potential is $O(1)$, say, and varies over $O(1)$ length scales, there will be an electron-charge boundary layer of thickness $O(\delta)=O(h)$ wherein $q=O(1)$. Hence, away from the gap we have to leading order the standard local conditions connecting the bulk vacuum potential and bulk metal potential.  Moreover, assuming, subject to \textit{a posteriori} confirmation, that  $\epsilon$ retains the scaling \eqref{eps scaling}, the outer metal-bulk potential is, as in \S\S\ref{ssec:dimerlocal}, uniform to leading order. In the vicinity of the gap, however, the situation is quite different. We still expect an electron-charge boundary layer of thickness $O(\delta)$, narrow relative to the pole domain studied in \S\S\ref{ssec:dimerlocal}; however, the comparably strong $O(1/h)$ transverse field in the gap implies through the no-flux boundary condition [third of \eqref{nondim bc}] that $q=O(1/h)$ in the boundary layer segment interacting with the gap; this increase in charge density results in a reshuffle of the asymptotics and consequently in the promotion of nonlocality to leading order.

We begin as in \S\S\ref{ssec:dimerlocal} by postulating a gap potential in the form $\varphi\sim \Phi+ O(h^{-1/2}) $, where $\Phi=-E(X)Y$. In describing the electron-charge layer adjacent to the gap region, we continue to employ the gap coordinates $X,Y$, with now $Y> H(X)+O(h)$. The scaling arguments given above suggest the boundary-layer expansions
\begin{equation}
q\sim h^{-1}Q_{-1} + h^{-1/2}Q_{-1/2} + \cdots, \quad \chi\sim T_0+h^{1/2}T_{1/2}+\cdots,
\end{equation}
where, as in \S\S\ref{ssec:macro}, it is 
  convenient to consider $\chi$ instead of the potential $\bar\varphi$. 
According to \eqref{eps scaling},
\begin{equation}
\frac{\epsilon}{\epsilon-1}\sim 1+O(h^{1/2}), \quad \frac{1}{\epsilon-1}\sim -\frac{1}{\zeta}h^{1/2}+O(h).
\end{equation}
Taking $\tau=\delta/h$ as an $O(1)$ parameter, we require the leading-order governing equations, that follow from \eqref{chi eq}, together with that governing $T_{1/2}$,
\begin{equation}\label{nonlocal leading}
\tau^2\pd{^2Q_{-1}}{Y^2}=Q_{-1}, \quad \pd{^2T_0}{Y^2}=0,\quad \tau\pd{^2T_{1/2}}{Y^2}=-\frac{Q_{-1}}{\zeta}.
\end{equation}
Similarly, the leading-order interfacial conditions follow from \eqref{chi bc} as
\begin{equation}\label{leading nonlocal bcs}
\pd{Q_{-1}}{Y}=\frac{1}{\tau}E(X), \quad \pd{T_0}{Y}=0, \quad E(X)H(X)=\tau Q_{-1}-T_0 \quad \text{at} \quad Y=H(X),
\end{equation}
and we will also need the next-order condition
\begin{equation}\label{T12 bc}
\pd{T_{1/2}}{Y}=0 \quad \text{at} \quad Y=H(X).
\end{equation}
Note that curvature is absent in \eqref{T12 bc} since the relative errors in approximating the boundary as $Y\sim H(X)$ and the normal derivative as $\partial/\partial{n}\sim(1/h)\,\partial/\partial{Y}$ are both $O(h)$. Additional conditions arise from matching with the transversally wider pole region, where $q$ is exponentially small and the potential is expanded as $\bar\varphi\sim \bar\Phi(X,\bar{Y})+O(h^{1/2})$, $\bar\Phi(X,\bar{Y})$ satisfying Laplace's \eqref{cyl pole eq} and the matching condition at infinity.  First, the exponential order of $q$ implies that 
\begin{equation}\label{Q dec}
Q_{-1},Q_{-1/2},\ldots \quad \text{as} \quad Y\to\infty.
\end{equation}
Second, assuming the pole potential is regular as $\bar{Y}\to0$,  expanding the leading term $\bar\Phi$ and rewriting in boundary-layer coordinates $(X,Y)$ gives
\begin{equation}
\Phi(X,\bar{Y}=0)+h^{1/2}Y\pd{\bar\Phi}{\bar Y}\left(X,\bar{Y}=0\right)+\cdots,
\end{equation}
whereby matching rules imply the conditions
\begin{equation}\label{Matching T}
T_0 \sim \Phi(X,\bar{Y}=0) + o(1), \quad T_{1/2} \sim Y\pd{\bar\Phi}{\bar Y}(X,\bar{Y}=0) + o(Y), \quad \text{as} \quad Y\to\infty.
\end{equation}

Integrating \eqref{nonlocal leading} in conjunction with \eqref{leading nonlocal bcs} and \eqref{Q dec} we find
\begin{equation} \label{Q0T0}
Q_{-1}= -E(X)e^{-[Y-H(X)]/\tau}, \quad T_0=-E(X)[H(X)+\tau].
\end{equation}
Next, integrating the third equation of \eqref{nonlocal leading} using \eqref{Q0T0} and \eqref{T12 bc} gives the asymptotic behaviour 
\begin{equation}\label{T der match}
\pd{T_{1/2}}{Y}\sim  \frac{1}{\zeta}E(X) \quad \text{as} \quad Y\to\infty.
\end{equation}
This, together with \eqref{Matching T}, \eqref{Q0T0} and \eqref{T der match}, yields the effective conditions connecting the pole and gap regions,
\begin{equation}\label{nonlocal interface}
\bar\Phi=-[H(X)+\tau]E(X), \quad \pd{\bar\Phi}{\bar Y}= \frac{1}{\zeta} E(X), \quad \text{at} \quad \bar Y=0,
\end{equation}
that combine giving a modified mixed boundary condition governing the pole region [cf.~\eqref{cyl mixed}],
\begin{equation}\label{new mixed}
\zeta[H(X)+\tau]\pd{\bar\Phi}{\bar Y}+\bar\Phi=0 \quad \text{at} \quad \bar Y=0.
\end{equation}

Together with \eqref{cyl pole eq} and the matching at infinity, the effective condition \eqref{new mixed} defines a new problem governing the metal-pole potential $\bar\Phi$. Remarkably, this problem is identical to that which we had found in the local case, other than the addition of $\tau$ to $H(X)$ in \eqref{new mixed}. In dimensional terms, this corresponds to an effective widening of the gap by {$2\lambda_F$}, as intuitively argued in Ref.~\onlinecite{Schnitzer:16}; the ``local'' pole problem of \S\S\ref{ssec:dimerlocal} is recovered by making the transformations 
\begin{equation}\label{trans}
X\to \sqrt{1+\tau}X, \quad \bar Y\to \sqrt{1+\tau}\bar Y, \quad \zeta\to \zeta/\sqrt{1+\tau}. 
\end{equation}
As a consequence, and using \eqref{cyl alpha} and \eqref{eps scaling}, we find the nonlocal eigenvalues 
 as 
\begin{equation}\label{nonlocal renorm}
\epsilon_n \sim - \frac{1}{\sqrt{2(h+\delta)}}\frac{1}{1+n}, \quad n=0,1,2,\ldots,
\end{equation}
which is a renormalisation of the singular local-theory eigenfrequencies \cite{Schnitzer:16}. From \eqref{trans}, the eigenmodes in the nonlocal case too are renormalised variants of the local ones, see the gap-field distributions in Fig.~\ref{fig:modeseo}.

In Fig.~\ref{fig:nonlocal_freq}, the blue dash-dot line depicts the fundamental $(n=0)$ surface-plasmon frequency predicted by \eqref{nonlocal renorm}, together with \eqref{Drude} and \eqref{eps scaling}], in the case of a gold dimer. The parameter set ($a=10nm$, $\hbar\omega_p=3.3\,eV$, $\beta=0.0036c$, hence $\delta=0.0215$) is chosen in order to compare with the numerical data of Luo \textit{et al.} \cite{luo:13}, extracted here using a plot digitiser and added as symbols in Fig.~\ref{fig:nonlocal_freq}. Also shown is the exact local result \eqref{dimer exact} --- black line, the local near-contact asymptotics \eqref{cyl alpha} --- black dash-dot, the perturbation result \eqref{eps 1 d} derived in \S\ref{sec:arbitrary} for $h=O(1)$ --- blue dashed, and a uniformly valid approximation to be discussed in the next section --- blue line. It should be noted that the error in \eqref{nonlocal renorm} is expected to asymptotically vanish as $h$ \emph{and} $\delta$ become small; since $\delta$ is held constant in Fig.~\ref{fig:nonlocal_freq}, the small error persisting as $h\to0$ is not surprising.

\section{A ``uniform'' coarse-grained model}
\label{sec:macro}
\begin{figure}[b]
   \centering
   \includegraphics[scale=0.25]{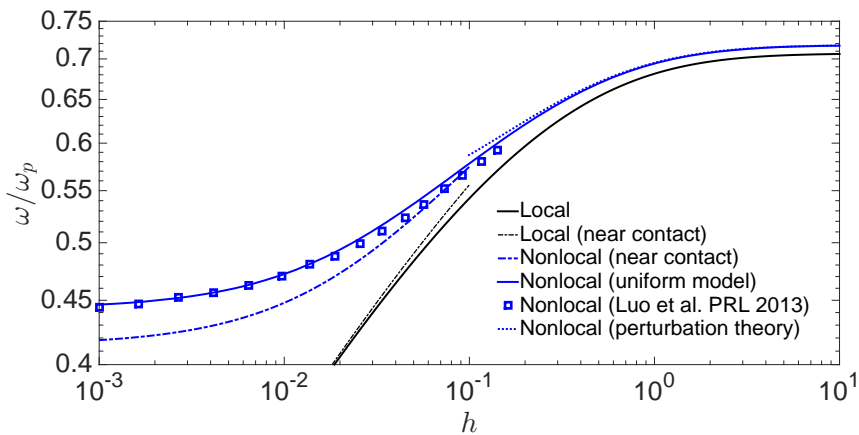} 
   \caption{Fundamental surface-plasmon bonding-mode of a gold circular cylindrical dimer ($a=10nm$, $\hbar\omega_p=3.3\,eV$, $\beta=0.0036c$, giving $\lambda= 0.215\,nm$ and $\delta=0.0215$). See text for details on the various data and approximations shown.} 
   \label{fig:nonlocal_freq}
\end{figure}

Our analysis  has proceeded in two complementary routes: In \S\ref{sec:arbitrary} we carried out an asymptotic coarse graining procedure valid for smooth single-scale metallic particles of otherwise arbitrary shape, while in \S\ref{sec:near} we performed an \emph{ab initio} asymptotic analysis of the near-contact limit of the cylindrical-dimer configuration (in which limit the former approach fails).  A ``uniform'' macroscale model, from which both approaches follow as special cases, can be obtained by solving Laplace's equation {for} the vacuum potential $\varphi$ and the metal-bulk potential $\bar\varphi$, in conjunction with the leading- and first-order effective {interfacial} conditions \eqref{eff 0} and \eqref{eff 1} derived in \S\ref{sec:arbitrary} fused together, i.e.
\begin{equation} \label{uniform}
\varphi-\bar\varphi \sim \delta \left(\frac{\epsilon-1}{\epsilon}\right)^{3/2}\pd{\varphi}{n}, \quad \pd{\varphi}{n}\sim\epsilon\pd{\bar\varphi}{n}.
\end{equation}
Evidently, for single-scale particles \eqref{uniform} yield blueshifted eigenfrequencies that are asymptotically identical to those predicted by the perturbation scheme of \S\ref{sec:arbitrary}. Less obviously, for multiple-scale configurations having narrow gaps, \eqref{uniform} again leads to the correct leading-order asymptotics as found in \S\ref{sec:near}. In the latter scenario, nonlocality is most appreciable in the gap region, where owing to the geometrically enhanced gap field the seemingly small potential discontinuity in \eqref{uniform} is promoted to leading order. Explicitly, as $h\to0$, we know that for bonding modes $|\epsilon|\gg1$ and hence $(\epsilon-1)/\epsilon\sim 1$; thus the second of \eqref{uniform} becomes $\varphi-\bar\varphi \sim \delta \partial{\varphi}/\partial{n}$. On the other hand, from the gap-scale analysis in  
\S\ref{sec:near} at the vacuum-metal interface $\varphi\sim -E(X)H(X)$ and $\partial\varphi/\partial n \sim h^{-1}E(X)$. We therefore see that \eqref{uniform} are to leading order asymptotically identical with the effective conditions \eqref{nonlocal interface}. In Fig.~\ref{fig:nonlocal_freq}, showing the fundamental bonding-mode eigenfrequency of a cylindrical dimer, the solid blue line depicts the  prediction obtained using \eqref{uniform}. At moderate $h$ the latter agrees with the perturbative result  \eqref{eps 1 d}. At small $h$ the new prediction agrees very well with the data of Ref.~\onlinecite{luo:13} and reasonably well with the renormalised near-contact asymptotics \eqref{nonlocal renorm}; as already explained in \S\ref{sec:near}, the finiteness of $\delta$ prevents the error in \eqref{nonlocal renorm} from vanishing as $h\to0$.  

\begin{figure}[t]%
       \centering
         \includegraphics[scale=0.28]{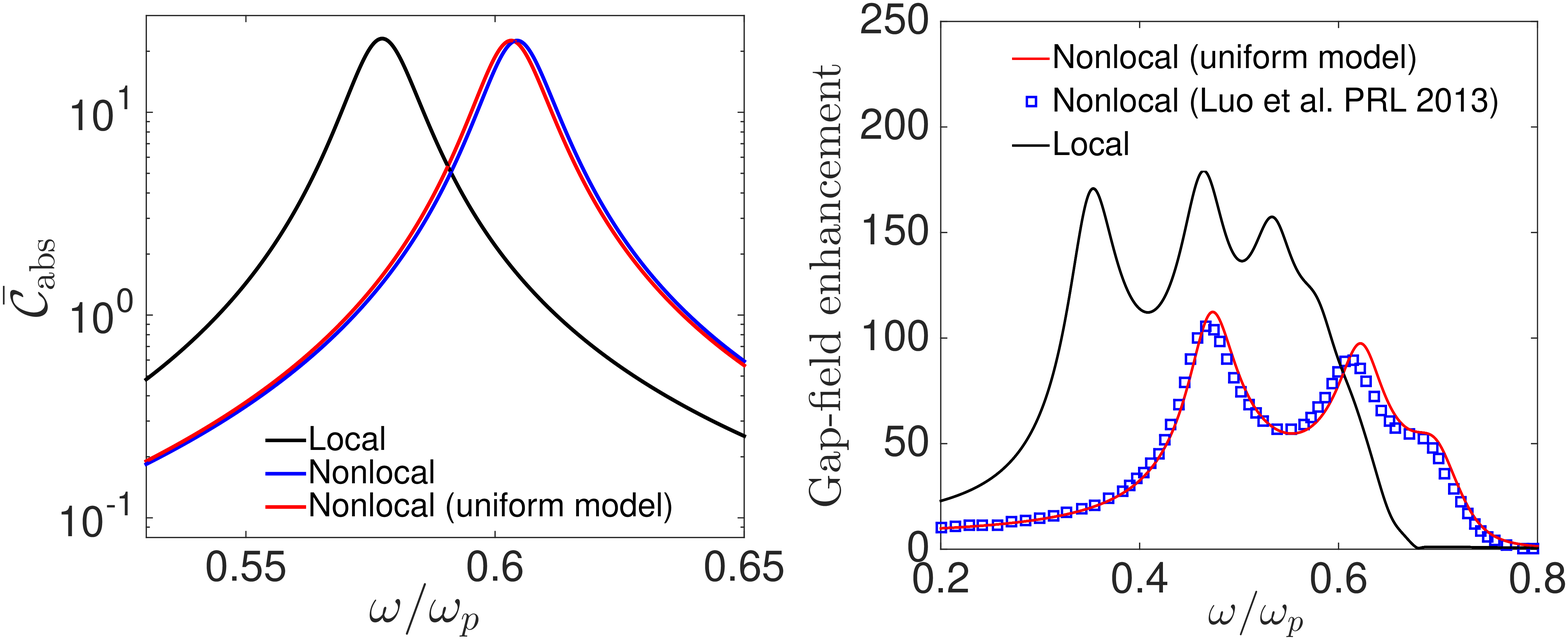} 
         \caption{Left: Normalised absorption cross section \eqref{normalised abs} of a nanometallic sphere under plane-wave illumination in the quasistatic limit --- same parameters as in Fig.~\ref{fig:example}. Black and blue lines respectively depict the local-theory prediction \eqref{mu sphere} and the nonlocal-theory prediction \eqref{mu sphere local}. Red line depicts the prediction \eqref{mu nonlocal} of the uniform local-analogue model \eqref{uniform}. Right: Field enhancement at the centre of a $0.2$\,nm-wide gap between a pair of gold cylinders of radius $a=10$\,nm ($h=0.01$), under plane-wave illumination with incident field polarised along the line of centres; same material parameters as in Fig.~\ref{fig:nonlocal_freq}. Black line --- solution of local model ; red line --- solution of the the uniform local-analogue model \eqref{uniform}; symbols --- numerical data extracted from Ref.~\onlinecite{luo:13} with a plot digitiser.} 
         \label{fig:uniform}%
\end{figure}

Importantly, the effective interfacial conditions \eqref{uniform} apply also to excitation problems. The coarse-graining perturbation scheme of \S\ref{sec:arbitrary}, in its original form, breaks down at frequencies close to resonant values as, under resonance conditions,  there are two small parameters: $\delta$, which together with geometry determines the resonance frequency in the absence of losses, and $\mathrm{Im}[\epsilon]/\mathrm{Re}[\epsilon]$, which governs the level of plasmonic enhancement at resonance. In the perturbation scheme of \S\ref{sec:arbitrary}, the leading-order description would resonate at the ``local'' eigenfrequencies rather than the blueshifted ones, thereby overpowering the $O(\delta)$ correction. A rigorous asymptotic procedure addressing the smallness of both parameters could be carried out be separately considering resonant and non-resonant  frequency regimes, or using the method of strained coordinates \cite{Hinch:91}. The fused conditions \eqref{uniform}, which uniformly capture the leading nonlocal effect at resonance, provide a simpler path ---  asymptotically sound, but admittedly not exploiting at all the smallness of $\mathrm{Im}[\epsilon]/\mathrm{Re}[\epsilon]$ towards simplification, as e.g.~in the local analysis of Ref.~\onlinecite{Schnitzer:16}.

As an example, consider the problem of plane-wave excitation of a metallic nanosphere, already solved in \S\ref{sec:sphere} by exact separation of variables of the hydrodynamic Drude model. Employing the simpler local-analogue model with interfacial conditions \eqref{uniform}, we readily find the dimensionless induced-dipole moment as [cf.~\eqref{mu sphere}]
\begin{equation}\label{mu nonlocal}
\mu \approx \frac{\epsilon-1-\epsilon \delta [(\epsilon-1)/\epsilon]^{3/2}}{\epsilon+2+2\epsilon\delta[(\epsilon-1)/\epsilon]^{3/2}}.
\end{equation}
The excellent agreement of \eqref{mu nonlocal} with \eqref{mu sphere} for small $\delta$ values is demonstrated in Fig.~\ref{fig:uniform} (left), where the normalised absorption cross section \eqref{normalised abs} of a metallic nanosphere under plane-wave illumination is plotted as a function of frequency. For a less trivial application of the uniform model, consider the scenario of plane-wave illumination of a metallic circular cylindrical dimer, with the incident electric field polarised along the line of centres. Fig.~\ref{fig:uniform} (right) compares as a function of frequency the field enhancement in the gap as predicted by the local model (black line), the uniform model (red line), and numerical data of Ref.~\onlinecite{luo:13}; material parameters and the gap-width parameter $h=0.01$ were chosen to match the latter simulation. The local solution was obtained in \cite{Klimov:14} by separation of variables in bipolar coordinates. For the uniform model, owing to the form of \eqref{uniform} separation of variables yields a system of algebraic equations which needs to be solved numerically (we omit the details). It should also be noted that the simulations in Ref.~\onlinecite{luo:13} are retarded, whereas the former two approaches are quasistatic.

\section{Recapitulation and concluding remarks} \label{sec:conclude}
The smallness of the {nonlocal screening length $\lambda_F$} relative to characteristic nanometer scales of plasmonic structures allows, through asymptotic analysis of the hydrodynamic Drude model, to extract in simple form the essential effects of nonlocality on surface-plasmon resonance. Thus, together with our previous study \cite{Schnitzer:16}, the present paper provides a general theory relating blueshifted surface-plasmon eigenfrequencies in the nonlocal case with eigenfrequencies and corresponding eigenmodes in the local approximation. 

In applying the theory it is important to distinguish between two cases. For nanometric structures characterised by a single length scale $a$ (or multiple length scales all $\gg\lambda_F$), the {blueshifts} are asymptotically small, $\Delta\omega/\omega_p=O(\lambda_F/a)$, and the perturbation $\Delta\omega=\omega_{\text{nonloc}}-\omega_{\text{loc}}$ is given by a quadrature over the corresponding unperturbed local eigenmode \eqref{eps 1}. In dimensional form, and using \eqref{om exp}, the perturbation formula reads
\begin{equation}\label{pert dimensional}
\frac{\Delta\omega}{\omega_p}\sim \frac{\lambda_F}{2}\frac{\oint |\bar\bE\bcdot \bn|^2\,\rm{d}A}{\int|\bar\bE|^2\,\rm{d}V}\sqrt{\frac{\omega_p^2}{\omega_{\text{loc}}^2}-1},
\end{equation}
where $\bar\bE$ denotes the local eigenmode field distribution within the metal domain, and the integrals in the numerator and denominator are respectively over the surface and volume of the metal domains. For a degenerate unperturbed  eigenfrequency the perturbation consists of a set of equations \eqref{new ev}, which, however, in most cases reduces to \eqref{pert dimensional}. We have successfully demonstrated the efficacy of this perturbation theory by deriving closed-form blueshift formulae for spheres \eqref{sphere correction}, cylinders \eqref{eps 1 cyl}, prolate spheroids \eqref{eps 1 s}, and circular cylindrical dimers \eqref{eps 1 d}. Whereas in these cases  the local eigenmodes are known analytically, it is clear that many more configurations can be studied via \eqref{pert dimensional} by harnessing standard numerical schemes and packages for calculating eigenmodes in the local approximation \cite{Mayergoyz:05,Hohenester:05,Hohenester:12}. 

The second case entails plasmonic structures characterised by multiple length scales, one of these being $O(\lambda_F)$. In such cases the perturbation formula \eqref{pert dimensional} may fail, as \eqref{dubious s} and \eqref{dimer break} show explicitly in the respective extreme cases of a slender elongated body of thickness {$O(\lambda_F)$} and a  nanowire dimer with an {$O(\lambda_F)$} gap width. The latter case falls under the general framework of Ref.~\onlinecite{Schnitzer:16}, where the eigenfrequencies of gap bonding modes in the near-contact limit are captured by a renormalisation of the singular redshift predicted by local theory. In this paper we corroborated this result by carrying out a detailed asymptotic analysis of the near-contact limit of a circular cylindrical dimer (radius $a$, gap width $d$). In agreement with Ref.~\onlinecite{Schnitzer:16}, the analysis showed the eigenfrequencies of the bonding modes to be 
\begin{equation}\label{renorm dim}
\frac{\omega_{\text{loc}}}{\omega_p} \sim \left(1+n\right)^{1/2}\left(\frac{d}{a}\right)^{1/4} , \quad \frac{\omega_{\text{nonloc}}}{\omega_p} \sim \left(1+n\right)^{1/2}\left(\frac{d+2\lambda_F}{a}\right)^{1/4},\quad n=0,1,2,\ldots,
\end{equation}
in the local and nonlocal levels of description, respectively. 

Both the perturbation formula for single-scale geometries \eqref{pert dimensional} and the near-contact renormalisation \eqref{renorm dim} hinge upon the narrowness of the electron-charge boundary layers forming at metal-vacuum interfaces relative to field and charge  variations along those interfaces. This separation of scales is evident for surface plasmons of single scale particles (excluding high-order modes). For bonding gap modes in the near-contact limit, however, this is less clear, since the near-field is highly confined to the vicinity of the gap. Nevertheless, as argued via scaling arguments in \cite{Schnitzer:16}, and demonstrated herein by the detailed analysis of \S\ref{sec:near}, the {$O(\lambda_F)$}-thick boundary layer remains thin relative to the transversal extent of the ``pole'' metal region wherein the field is confined. Indeed, the latter is $O[(da)^{1/2}]$ in the absence of nonlocality, and {$O[(d+2\lambda_F)^{1/2}a^{1/2}]$} in the strongly nonlocal limit {$d=O(\lambda_F)$}; hence, the separation of scales prevails all the way up to the theoretical redshift saturation. 

The above scale separation enables an asymptotic coarse graining procedure, where the physics of the narrow electron-charge layer are replaced by a set of effective ``local-analogue'' interfacial conditions. In Ref.~\onlinecite{Schnitzer:16}, we put forward such an effective coarse-grained ``local-analogue'' eigenvalue problem based on scaling arguments and an intuitive one-dimensional analysis. That effective problem consists of finding quasistatic fields $\bar\bE$ and $\bE$ that are solenoidal in the metal and vacuum domains, respectively, with $\bE$ attenuating at large distances, and  the fields satisfying the effective interfacial conditions
\begin{equation}\label{macro dim}
\varphi'-\bar\varphi' \sim {\lambda_F} \left(\frac{\epsilon-1}{\epsilon}\right)^{3/2}\pd{\varphi}{n}, \quad \pd{\varphi'}{n}\sim\epsilon\pd{\bar\varphi'}{n},
\end{equation}
where $\epsilon$ is the relative Drude permittivity \eqref{Drude}, and $\varphi'$ and $\bar\varphi'$ are electric potentials such that $\bE=\mathrm{Re}[-\bnabla\varphi' \exp(-i\omega t)]$ and $\bar\bE=\mathrm{Re}[-\bnabla\bar\varphi' \exp(-i\omega t)]$. Here we have substantiated this model, clarifying its validity and asymptotic accuracy under various circumstances, including to plasmonic excitation problems. In particular, using the method of matched asymptotic expansions, we systematically derived the above model for the surface-plasmon modes of single-scale particles. In that case, the relative effect of nonlocality is {$O(\lambda_F/a)$} while the relative error of the model is {$O(\lambda_F^2/a^2)$}. We further showed that the above model correctly furnishes the leading-order description of gap bonding modes in the near-contact limit, where nonlocality is no longer a small perturbation. Thus, for a dimer configuration the effective conditions \eqref{macro dim} encapsulate the leading-order effect of nonlocality across all gap widths.

\section*{Acknowledgements}
RVC and SM acknowledge funding from the Engineering and Physical Sciences Research Council via Programme grant EP/L024926/1. SM acknowledges further support from the Royal Society and from the Lee-Lucas Chair that he holds.

\bibliography{refs2.bib}

\end{document}